\documentclass[10pt,journal,compsoc]{IEEEtran}
\usepackage[T1]{fontenc}
\usepackage[linesnumbered,vlined,ruled,commentsnumbered]{algorithm2e}
\usepackage{bm}
\usepackage{hyperref}
\usepackage{latexsym, graphicx, textcomp}
\usepackage{epsfig,upgreek}
\usepackage{amsfonts, amssymb, amsthm, makecell, enumerate}
\usepackage{float,color,dblfloatfix}
\usepackage{fancyhdr}
\usepackage{amsmath}
\usepackage{autobreak}
\usepackage{mathrsfs}
\usepackage{pifont}
\usepackage{amssymb}
\usepackage{graphicx}
 \usepackage{epsfig}
\usepackage{multicol}
\usepackage{cases}
\usepackage{enumerate}
\usepackage{cite}
\usepackage{booktabs}
\usepackage{setspace}
\usepackage{subfigure}
\usepackage{pifont}
\usepackage[hang,flushmargin]{footmisc}
\usepackage{setspace}
\usepackage{multirow} 
\usepackage{cuted}
\usepackage{multicol}
\usepackage{ragged2e}
\usepackage{bbm}
\usepackage{extarrows}
\usepackage{xcolor}

\def\footnoterule{
\kern-5pt
\hbox to \columnwidth{\vrule width 0.618\columnwidth height 0.4pt\hfill}
\kern4.6pt}

\def\ge{\geqslant}
\def\geq{\geqslant}
\def\le{\leqslant}
\def\leq{\leqslant}

\graphicspath{{figures/}}

\newtheorem{pf*}{Proof}

\ifCLASSINFOpdf

\else

\fi

\allowdisplaybreaks
\begin{document}
\title{ Long-Term or Temporary?\\Hybrid Worker Recruitment for \\ Mobile Crowd Sensing and Computing}

\author{Minghui Liwang,~\IEEEmembership{Member, IEEE,} Zhibin Gao,~\IEEEmembership{Member, IEEE,} Seyyedali Hosseinalipour, ~\IEEEmembership{Member, IEEE,} Zhipeng Cheng,~\IEEEmembership{Student Member, IEEE,} 
Xianbin Wang,~\IEEEmembership{Fellow, IEEE,} and Zhenzhen Jiao,~\IEEEmembership{Member, IEEE} 
\thanks{Minghui Liwang (minghuilw@xmu.edu.cn) is with the Department of Information and Communication Engineering, Xiamen University, Xiamen, China. Zhibin Gao (gaozhibin@jmu.edu.cn, Corresponding author) is with the Navigation Institute, Jimei University, Xiamen, China.
Zhipeng Cheng (chengzp\_x@163.com) is with the School of Future Science and Engineering, Soochow University, Suzhou, China. Seyyedali Hosseinalipour (hosseina@
purdue.edu) is with the School of Electrical and Computer Engineering, Purdue University, West Lafayette, Indiana 47906 USA. Xianbin Wang (xianbin.wang@uwo.ca) is with the Department of Electrical and Computer Engineering, Western University, London, Ontario N6A 5B9, Canada. Zhenzhen Jiao (jiaozhenzhen@teleinfo.cn) is with the iF-Labs, Beijing Teleinfo Technology Company,
Ltd., CAICT, Beijing 100734, China.

\noindent

}}

\IEEEtitleabstractindextext{
\begin{abstract}
\justifying
This paper investigates a novel hybrid worker recruitment problem where the mobile crowd sensing and computing (MCSC) platform employs workers to serve MCSC tasks with diverse quality requirements and budget constraints, under uncertainties in workers’ participation and their local workloads. We propose a hybrid worker recruitment framework consisting of offline and online trading modes. The former enables the platform to overbook long-term workers (services) to cope with dynamic service supply via signing contracts in advance, which is formulated as 0-1 integer linear programming (ILP) with probabilistic constraints of service quality and budget. Besides, motivated by the existing uncertainties which may render long-term workers fail to meet the service quality requirement of each task, we augment our methodology with an online temporary worker recruitment scheme as a backup Plan B to support seamless service provisioning for MCSC tasks, which also represents a 0-1 ILP problem. To tackle these problems which are proved to be NP-hard, we develop three algorithms, namely, \textit{i)} exhaustive searching, \textit{ii)} unique index-based stochastic searching with risk-aware filter constraint, \textit{iii)} geometric programming-based successive convex algorithm, which achieve the optimal or sub-optimal solutions. Experimental results demonstrate our effectiveness in terms of service quality, time efficiency, etc. 

\end{abstract}

}

\maketitle

\IEEEpeerreviewmaketitle

\section{Introduction}

\IEEEPARstart{T}{he} past decade has witnessed an enormous increase in the number of intelligent Internet of Things (IoT) devices embedded with powerful processors and sensors, e.g., camera, GPS, gyroscope. This development has led to the rise of a wide range of innovative data- and computation-intensive applications (e.g., traffic information gathering, urban WiFi characterization, weather forecasting, and social services)~\cite{1} that aim to process the distributively collected data at the network edge via modern computation techniques~\cite{2,3,4,5}. Traditional techniques such as sensor networks have been separately implemented over real-world networks, which, however, may impose high installation cost, suffer from limited spatial coverage, and face difficulties in integrating the distributed mobile computing/storage resources~\cite{6}. 

To this end, \textit{Mobile Crowd Sensing and Computing} (MCSC) at network edge has been introduced~\cite{6,7}. MCSC represents a novel platform that enables ordinary users to contribute sensed/generated data from their mobile devices, while aggregating and processing heterogeneous crowd sensed data at the network edge for intelligent service provisioning~\cite{7,8}, e.g., intelligent urban traffic management. In MCSC platform, IoT devices can act as mobile device clouds (MDCs)~\cite{9,10} equipped with on-board processors and available resources, that pre-process their collected data locally to filtrate useless information, which leads to data redundancy reduction while alleviating the heavy burden of device-to-edge communication and edge server computation processes. 

MCSC generally constitutes a three-tiered hierarchical network: \textit{i)} an edge-based MCSC platform as a centralized controller and data processing center (service requestor), \textit{ii)} multiple Point-of-Interests (PoIs, a PoI represents a certain region where the MCSC platform is interested in its information and data) coordinated by the edge server, and \textit{iii)} heterogeneous mobile smart devices (service providers) which are referred to \textit{workers}. Each PoI can announce various tasks with different service quality requirements (e.g., data quality, pre-processing delay, etc.), while a set of workers around the corresponding PoI are recruited to offer sensing and computing services\cite{11}. Nevertheless, pre-processing a massive amount of collected data on IoT devices and transmitting the corresponding results to the MCSC platform may require individual efforts~\cite{11}, bringing crucial issues, e.g., extra computation and communication costs. Also, the selfishness of participants (especially workers)~\cite{12} may hinder a  smooth service provisioning procedure, which calls for proper monetary incentives~\cite{6}. These challenges advocate studying MCSC as a service trading market, where the MCSC platform pays commission to each worker according to its contribution to a specific task. To put it succinctly, in this paper, we study trading-empowered MCSC, where during a trading, each recruited worker pre-processes its collected raw data which is required for the successful execution of an MCSC task locally, and transmits the corresponding results to the platform, while receiving a certain payment. 

\subsection{Motivations}
\noindent
Existing studies on MCSC have been focused on designing online and offline trading mechanisms for worker recruitment. Online MCSC considers the current information associated with workers and tasks during each practical trading (e.g., information associated with a network snapshot), to conduct online decision-making (e.g., worker-task assignment)~\cite{13}. Although online trading can capture the current task/worker/network conditions and achieves relatively accurate decisions, it faces major difficulties to cope with the random and dynamic nature of mobile networks. For example, mobile devices may choose to participate or to be absent from a trading due to \textit{i)} uncertain mobility, and \textit{ii)} uncertain dynamic local workloads, which results in dynamic and unwarranted service supply. Furthermore, time-varying wireless network conditions impose uncertainties on the availability of sensing and computing services, e.g., a large data transmission delay can be incurred by poor wireless channel quality between a worker and the gateway of the MCSC platform. Major challenges associated with online MCSC caused by the uncertainties are detailed below: 

\noindent
$\bullet$ \textit{Extra latency incurred by online decision-making}: Temporal variations of system states (e.g., time-varying channel quality between each worker and the gateway of MCSC platform) require MCSC platform and workers (together referred to participants) to spend excessive amount of time to analyze the current tasks/workers/networks-related information to reach the final agreement during every trading, which reduces the amount of time that can be used for practical sensing and computing. Specifically, since an online decision only works for the current trading, an accumulated latency will be incurred in the long-term upon considering a large number of online trading. 

\noindent
$\bullet$ \textit{Extra energy consumption incurred by online decision-making}: A long online decision-making procedure can lead to excessive energy consumption, and subsequently a considerable carbon emission and air pollution in long-term~\cite{14}. This factor can cause both performance degradation and undesired trading experience, especially in wireless networks with energy-limited mobile devices~\cite{2,15}. 

\noindent
$\bullet$ \textit{Unexpected trading failures}: Online trading may lead to undesired trading failures, and low quality of experience for participants. A typical example is online auctions, where a limited number of winners (winning workers) obtain the eventual employment contracts, while no compensation is offered to the losers (workers who fail in a trading) which also have spent time/energy on bidding/negotiating/waiting. Trading failures can significantly impact the trading experience of participants especially for workers, making them no longer willing to offer services. 

We are thus motivated to develop cost-effective and time-efficient service provisioning approaches for MCSC in wireless networks, given the above-mentioned drawbacks of online trading. Offline (in-advance) decision-making offers a feasible solution with less impact on the real-time service delivery~\cite{16,17,18}. Nevertheless, most of studies on offline trading design for MCSC generally consider known prior information (e.g., prior knowledge of the workers’ locations/resources), which can be impractical to obtain in real-world networks~\cite{13,19}. 

Given the advantages of online and offline trading modes, this paper introduces a novel hybrid worker recruitment mechanism integrating both \textit{offline trading} that relies on historical trading statistics~\cite{18}, and \textit{online trading} that considers the current worker/task/network conditions. 
In our developed platform, the offline trading mode determines \textit{long-term workers} for each task by signing risk-aware contracts in advance, taking into account the historical characteristics of workers/tasks/networks. Long-term workers will offer sensing and computing services to the corresponding tasks during each trading while receiving pre-negotiated payments, without any further negotiation/bargain/bidding with the MCSC platform. This strategy significantly reduces the overhead (e.g., time and energy) of online MCSC decision-making. Specifically, our considered offline trading mode enables the MCSC platform to \textit{overbook}~\cite{21,22} long-term services from workers, which implies that, the overall service quality promised in pre-signed contracts for a task can be higher than that of its actual demand, given the dynamic service supply. 
Overbooking helps offering satisfactory quality of service when some long-term workers cannot offer services at some particular trading times, or opt out of a trading~\cite{16}, e.g., when they are located outside of the corresponding PoI due to their mobility. 
Although it is important to recruit enough workers for each task, the number of workers is usually constrained by each task’s budget. Thus, the overbooking should be carefully investigated to alleviate the risk of going over budget. 

Although overbooking is allowed, the overall service quality offered by long-term workers for a task may be unsatisfying owing to the uncertain workers' participation and fluctuant quality of service. 
To this end, we develop an online trading mode as an alternative Plan B to achieve better service quality for MCSC tasks. In the online trading mode, the platform selects \textit{temporary workers}\footnote{Namely, workers those who have not been selected as long-term workers are called “temporary workers” in this paper.} during each trading when the service requirement (e.g., data quality) of some tasks are not met by long-term workers.

\subsection{Relevant Investigations}
\noindent
We review related works devoted to the worker recruitment problem in mobile crowd sensing (MCS)\footnote{This paper considers MCSC, where the raw sensing data can be pre-processed on mobile devices first and then send the useful information to the platform, avoids data redundancy while alleviating heavy burden on both backhaul links and edge/cloud servers~\cite{9,10}. However, most existing works have put emphasis on MCS, and neglected the computation and processing aspects of the system to some extent. Thus, MCS is highly relevant to our topic.}, where worker recruitment mechanism is designed while assuming that workers’ information (e.g., sensing qualities) are known in advance \cite{12,23,24,25}, which, is impractical in some real-world scenarios. 

Accordingly, a few existing works have been dedicated into scenarios with partially unknown information, e.g., unknown sensing qualities and locations of workers. Such studies can either be related to economic behaviors~\cite{11,13,19,26,27,28,43,44,46,47,48} or not~\cite{41,42,45}. Regarding service trading markets,
\textit{Wang et al.} in~\cite{13} considered location-aware and location diversity-based dynamic crowd sensing systems with moving workers and stochastic arrival tasks.
\textit{née Müller et al.}~\cite{26} introduced a context-aware hierarchical online learning algorithm to maximize the performance of MCS under unknown workers. In~\cite{27}, \textit{Xiao et al.} investigated the worker recruitment problem with unknown service qualities, aiming to maximize the total sensing quality under a limited budget, while ensuring workers’ truthfulness and individual rationality. Similarly, \textit{Gao et al.} in~\cite{19} assumed that workers’ sensing qualities and costs are unknown a priori, and focused on continuous sensing tasks. Assuming that sensing quality needs to be protected from disclosure, \textit{Zhao et al.} in~\cite{28} modeled the worker recruitment under unknown service quality as a differentially private multi-armed bandit game.  
\textit{Wu et al.}\cite{43} investigated a utility-based sensing task decomposition
and subcontract algorithm, by establishing direct collaboration between mobile nodes (namely, workers).
In \cite{44}, \textit{Gao et al.} utilized UAVs to optimize the sensing coverage and data quality, by jointly considering the optimization of task allocation and trajectory scheduling. In \cite{46}, \textit{Zhang et al.} focused on two task types, namely, long-duration (e.g., time-sensitive) and short-duration (e.g., data-intensive) tasks and proposed a budget re-distribution algorithm for worker recruitment in cloud-aided edge networks.~\textit{Li et al.} in \cite{47} studied a POI-tagging App-assisted incentive mechanism (PTASIM), which explores the cooperation with POI-tagging App for mobile edge crowdsensing. Specifically, PTASIM requests App to tag some edges to be POI, and further guides App users to perform tasks at the corresponding location. In \cite{48}, \textit{Zhang et al.} 
proposed a mobility prediction-based multi-task allocation method, by using workers’ historical trajectories more comprehensively via fuzzy logic system to achieve better predictions, while designing a global heuristic searching algorithm to optimize the overall task completion rate. Among existing works, the most relevant work is~\cite{11}, where \textit{Xu et al.} designed a pricing scheme which concerns with: the payment offered to the workers, based on their reputations (obtained from historical performance) and their actual performance (during practical trading). Nevertheless,  although the above-mentioned studies have investigated the worker recruitment problem from various perspectives, our paper introduces a different angle, as discussed below. 

\noindent
$\bullet$ \textit{A resource overbooking-empowered hybrid trading mode}: To the best of our knowledge, our study is among the first to introduce a hybrid trading mode by integrating both offline and online trading, which is more applicable in real-world networks with dynamics. For example, in our considered market, a task can be completed with the help of both pre-signed long-term workers and temporary workers (as backup services). More importantly, this paper encourages tasks to overbook services in coping with the uncertain resource supply, representing a novel research highlight to the literature. 
Due to overbooking, we further incorporate soft budget constraints into the mechanism design, where the budget of each task can slightly fluctuate, which is more practical as compared to hard budget assumption that does not allow any changes (e.g,\cite{12}). 

\noindent
$\bullet$ \textit{Uncertain factors}: We consider various uncertainties in the MCSC network. For example, unpredictable workers’ participation due their mobility. Also, we analyze the service quality under a more fine-grained representation, in which the local workload of each worker is considered (which has rarely been studied in literature). Interestingly, in our considered scenario, each worker can have his own tasks that require multi-dimensional on-board resources, e.g., storage and computing, which can impact the service quality offered to MCSC tasks. 

\noindent
$\bullet$ \textit{Risk evaluation and control}: Risks caused by uncertainties have been neglected in many studies. In our market, since the uncertainties may cause economic losses (e.g., over budget) or unsatisfying services, we evaluate and control possible risks that both workers and the platform may face to protect service qualities and workers' profits.

\subsection{Key Challenges}
\noindent
We aim to solve the worker recruitment problem for MCSC in wireless networks via considering two stages: \textit{i)} long-term worker determination (offline mode); and \textit{ii)} temporary worker recruitment during each trading (online mode). 
We are facing three key challenges during the first stage: 

\noindent
\textit{i) How to determine feasible long-term workers for each task?} This is challenging since the service qualities and the participation of workers are unknown prior to each practical trading. 

\noindent
\textit{ii) How to determine long-term contract terms, such as the payment to a hired worker, and the quality of service that this worker should provide?} This is crucial since, on the one hand, tasks can have different budgets and service demands; while, on the other hand, a small payment may lead to negative worker’s utility. This paper considers two levels of quality assurance for each task, namely, \textit{hard} and \textit{soft}, to identify the sensing and computing service quality provided by workers.

\noindent
\textit{iii) How to decide the overbooking rate for each task under the service quality requirement and the budget constraint?} This is an important aspect of the design, since a large overbooking rate can lead to the case where the total payment to long-term workers exceeds the task budget, while a small rate may result in unsatisfying service quality. Although temporary workers can be exploited, online decision-making to recruit them imposes network costs, e.g., latency and energy.

For the second stage (i.e., temporary worker recruitment), our key research question is: \textit{How to determine temporary workers for a task when the corresponding long-term workers fail to meet its service quality requirements?}

In summary, we solve the fundamental problem
of \textit{Joint task-to-worker association, as well as payment and service level determination in both online and offline trading modes in MCSC networks.}

\subsection{Outlines and Summary of Contributions}
\noindent
Our considered scenario involves: \textit{i)} an MCSC platform as a service requestor (e.g., edge server), \textit{ii)} PoIs that can generate tasks with service quality requirements and certain budgets, and \textit{iii)} multiple workers (e.g., mobile devices) with different capabilities. Also, significant uncertainties are modeled and incorporated into our methodology to better capture the random and unpredictable nature of MCSC networks. For each worker, two levels of service quality are considered: hard, and soft quality. The former implies that the worker assigns the highest priority to the scheduled MCSC task, to guarantee a hard service quality assurance by charging a high price\footnote{In this paper, the word “price” indicates the “payment” to each worker, which are interchangeable with each other.}. The latter refers to a fluctuating service quality, where the worker will first execute its local workload and then processes the assigned MCSC task, under a lower price.
To the best of our knowledge, this work is among the first to study the unknown worker recruitment problem for MCSC networks via considering hybrid trading mode, overbooking, and risk evaluation. Our major contributions are summarized below:

\noindent
$\bullet$ We introduce a novel hybrid worker recruitment methodology for MCSC service provisioning via unifying both offline and online trading modes, where uncertain workers’ participation and fluctuant workers' local workloads are considered to capture the random and unpredictable nature of the network. 

\noindent
$\bullet$ The offline worker recruitment mode encourages the MCSC platform to overbook services from long-term workers for each task, in coping with dynamic service supplies, by signing employment contracts with promised payment and service quality level in advance, via analyzing historical statistics associated with uncertainties.
Motivated by the existence of uncertainties which may render long-term workers with contracts fail to meet the service quality requirement of each task, we further complement our methodology with an online temporary worker recruitment scheme as an efficient backup.

\noindent
$\bullet$ The proposed long-term worker recruitment problem is formulated as an optimization problem that maximizes the expected utility of the platform, under acceptable risks of both workers and tasks (which are probabilistic constraints), representing a 0-1 integer linear programming (ILP) problem with NP-hardness. We then design three algorithms to tackle the 0-1 ILP problem: \textit{i)} exhaustive searching, \textit{ii)} unique index-based stochastic searching with risk-aware filter constraint, and \textit{iii)} geometric programming-based successive convex algorithm, obtaining either optimal (with high computational complexity) or sub-optimal (with low computational complexity) solutions. Similar with the above three algorithms associated with offline mode, the optimization of online mode aims to select temporary workers for maximizing the overall practical utility of MCSC tasks those with unsatisfying qualities, via analyzing the current network/task/worker conditions. 
   
\noindent
$\bullet$ Comprehensive simulations relying on both numerical data and real-world dataset demonstrate that our proposed overbooking-enabled and risk-aware hybrid worker recruitment mechanism for MCSC tasks outperforms baseline methods from different perspectives, e.g., long-term service quality, time efficiency (running time), etc. 

The rest of this paper is organized as follows. In Section~2, we provide an overview of MCSC networks and introduce our modeling. Long-term worker determination and contract design problems, as well as temporary worker recruitment problem are proposed and analyzed in Section~3. Experimental results are carried out in Section~4, before drawing the conclusion in Section~5.

\section{Overview and System Model}
\subsection{Overview}

We consider an MCSC network consisting of multiple workers gathered via set $\mathbb{W}=\{w_i|i\in \{1,2,...,|\mathbb{W}|\}\}$ that can be recruited to collect and compute data for multiple tasks of interest to various PoIs~\cite{29,30} collected via the set $\mathbb{O}=\{{PoI}_j|j\in \{1,2,...,|\mathbb{O}|\}\}$. Each ${PoI}_j$ periodically\footnote{In this paper, we mainly consider periodic sensing tasks, which are also common in real-world networks. For example, a PoI may be interested in the traffic and pedestrian data associated with a certain road intersection during rush hours (e.g., 7:00am-9:00am and 17:00am-19:00am, every workday), and thus periodically generates MCSC tasks every 10 minutes. Correspondingly, there will be a trading in each time slot 7:00am, 7:10am, 7:20am, etc., and contracts can be signed among MCSC platform and workers especially for rush hours. Note that our proposed methodology can also be applied for different task arrival rates. For example, considering two task types, where tasks of type 1 generate every 5 minutes, and tasks of type 2 generate every 10 minutes. Then, our considered hybrid market can allow different contracts for different task types, which can be implemented independently.} generates a certain number of sensing tasks expressed as $\mathbb{S}_j=\{\bm{s_{j,k}}|k\in \{1,2,...,|\mathbb{S}_j|\}\}$, for the execution of recruited workers through a sequence of trading. During each trading, a worker offers service (namely, contribute and pre-process data) to a task $\bm{s_{j,k}}$ if it stays within the relevant $PoI_j$’s region \cite{11,23,32} while charging a certain price. We consider two modes for worker recruitment, as detailed below. 


\noindent
\textit{\textbf{i) Offline long-term worker recruitment}}: This procedure happens in prior to future trading, where the MCSC platform recruits workers for each task via signing long-term contracts\footnote{Considering and optimizing the expiration date of each long-term contract is out of scope in this paper. For example, the MCSC platform can terminate the contracts after a certain period of time and update historical statistics to achieve better worker recruitment solutions.} in advance, which will be fulfilled accordingly during each trading without any further negotiations. This mode encourages each task to overbook services from workers to cope with the underlying uncertainties in the system. The terms of contract involves a task (namely, a specific task associated with a specific PoI), a worker, the relevant service quality assurance (hard or soft, which will be introduced in Sec. 2.3), as well as the payment. Workers who have signed contracts with the MCSC platform regarding the tasks of ${PoI}_j$, are referred to \textit{long-term workers} of ${PoI}_j$.

\noindent
\textit{\textbf{ii) Online temporary worker recruitment}}: This procedure occurs during each trading. Due to the dynamics and uncertainties associated with MCSC networks, a task may suffer from the risk of receiving unsatisfying service quality\footnote{Although overbooking offers a higher probability of long-term workers' participation, it still suffers from the risk of insufficient contractual workers and low service quality involved in a trading, mainly caused by uncertainties $\alpha_{i,j}$ and $\beta_i$,  introduced by Sec. 2.3.}. For example, long-term workers who happen to move outside of the target PoI’s region will fail to provide services to their assigned tasks. At this time, MCSC platform can recruit \textit{temporary workers} without pre-signed contracts for the tasks with unsatisfying service quality, under an online trading mode. Notably, online worker recruitment procedure relies on analyzing the current task/worker/network conditions, which may incur extra latency and energy consumption. 
\begin{figure*}[!t]
\centering
\includegraphics[width=1\linewidth]{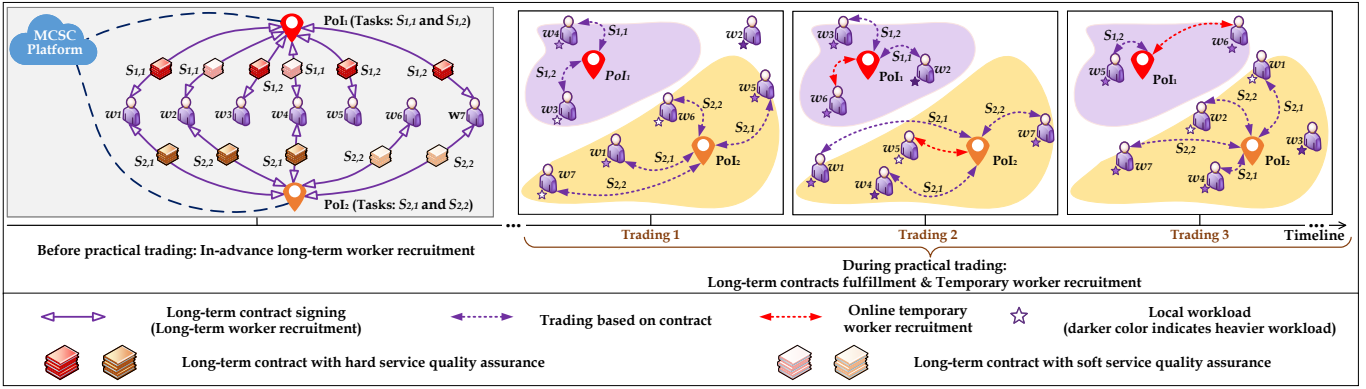}
\caption{Framework and timeline associated with the proposed hybrid worker recruitment for MCSC in wireless networks. The purple and yellow shadows describe the region of PoIs, while the dark-colored contract indicates the hard service quality assurance and the light-colored one means the soft quality assurance. Besides, the star aside each worker denotes the relevant local workload, where the darker color indicates a heavier local workload that may require more resources (e.g., storage, computing resources, etc.). }
\label{fig1}
\end{figure*}

Fig. 1 illustrates the timeline for our proposed hybrid worker recruitment mechanism, which is segmented into two distinct phases: \textit{i)} before trading, and \textit{ii)} during trading. During the former phase, the MCSC platform engages workers by establishing long-term contracts. Subsequently, in the trading phase, both long-term workers and the MCSC platform execute their commitments as outlined in the previously agreed contracts, with the potential addition of temporary workers to supplement the workforce as needed. Also, several practical trading examples are illustrated in Fig. 1, e.g., in trading 1, long-term worker $w_2$ fails to serve tasks $\bm{s_{1,1}}$ and $\bm{s_{2,2}}$ since it moves outside of both the two associated PoIs’ coverage zone.

\subsection{Task Modeling}
\noindent
We denote task $\bm{s_{j,k}}$ belonging to ${PoI}_j$ as a tuple $\bm{s_{j,k}}=\left(d_{j,k}^{B},d_{j,k}^{Q}\right)$, where $d_{j,k}^{B}$ denotes the tolerable budget of the task, which affects the number of employed workers. Specifically, the budget can be distributed among the two types of workers. For instance, in each practical trading, a task should first pay its long-term workers (who have attended) for the promised prices, while the remaining budget can be utilized for recruiting temporary workers in improving its received service quality.
Besides, $d_{j,k}^{Q}$ represents the desired service quality~\cite{12}, which incorporates the following factors: \textit{i)} data quality, reliability and accuracy, e.g., high and low-resolution image data can lead to different data qualities; \textit{ii)} sensing duration, e.g., some workers may need to travel around a certain region for data collection \cite{11,12,33}; \textit{iii)} time for data analyzing (computing), e.g., a worker may have to pre-process the collected raw data locally, to reduce possible data redundancy and heavy computation burden on the MCSC platform; \textit{iv)} time for uploading data to the MCSC platform through the wireless medium. 

\subsection{Worker Modeling}
We assume that each worker $w_i\in\mathbb{W}$ can contribute data and computing service to one task within a PoI region during a trading. To better capture the unpredictable and random nature of MCSC networks, we consider two key uncertainties associated with each worker $w_i\in\mathbb{W}$. 

\noindent
$\bullet$ \textit{\textbf{Uncertain participation.}} This uncertainty describes whether a worker $w_i$ is located inside or outside from ${PoI}_j$ during a trading process, due to its mobility (in the latter of which it cannot offer sensing and computing service to ${PoI}_j$). For worker $w_i\in \mathbb{W}$, let random variable $\alpha_i$ describe its attendance in different PoIs' regions that follows a discrete distribution\footnote{Note that in our problem formulation (see Section 3), to facilitate the analysis, we mainly consider one PoI, where the corresponding distribution of $w_i$ can reasonably be modeled as a Bernoulli distribution independent of time. Moreover, we use different values of $a_i$ to describe various behaviors of workers.}
denoted by $\alpha_i\sim \textbf{D}\left\{\left(0,1,...,j,...,|\mathbb{O}|\right), \left(1-\sum_{j=1}^{j=|\mathbb{O}|} a_{i,j}, a_{i,1},...a_{i,j},...,a_{i,|\mathbb{O}|}\right)\right\}$, where $0\le \sum_{j=1}^{j=|\mathbb{O}|}a_{i,j} \le 1$. Specifically, $\alpha_i=j$ with probability $a_{i,j}$ captures an event in which worker $w_i$ is within ${PoI}_j$’s coverage; while ${\alpha_i=0}$  with probability $1-\sum_{j=1}^{j=|\mathbb{O}|}a_{i,j}$ means that worker $w_i$ is absent from the trading (“no show”)~\cite{16} since it is located outside of all PoIs' regions and can not offer any service. For notational simplicity, let $\mathbb{A}=\{\alpha_1,...,\alpha_i,...,\alpha_{|\mathbb{W}|}\}$ where $\alpha_i\in\mathbb{A}$ are independent from each other. 

\noindent
$\bullet$ \textit{\textbf{Uncertain local workload.}} In practice, workers may also need to handle their local tasks, which can impact the service quality on performing MCSC tasks. For example, a worker’s local tasks may spend a certain period of time for waiting due to the completion of an assigned MCSC task (e.g., wait for the release of occupied computing/storage resources) \cite{17}. The local workload of each worker $w_i\in\mathbb{W}$ is modeled by random variable $\beta_i$ where $b_{w_i}^{-}\le \beta_i\le b_{w_i}^{+}$, following a normal distribution with mean $\mu_{w_i}$ and variance ${(\sigma_{w_i})}^2$: $\beta_i$ obeys a truncated normal distribution~\cite{35} $\beta_i\sim \textbf{N}\left(\mu_{w_i},{(\sigma_{w_i})}^2, b_{w_i}^{-}, b_{w_i}^{+}\right)$.  A large value of ${\beta_i}$ indicates a heavy local workload that $w_i$ needs to handle. For analytical simplicity, let $\mathbb{B}=\{\beta_1,...,\beta_i,...,\beta_{|\mathbb{W}|}\}$ where all $\beta_i\in\mathbb{B}$ are independent from each other. 
Note that workers in the hybrid market are truthful, namely, they will not misreport their local workload (e.g., pretend to have a high workload) in either offline or online modes, due to the following reasons. First, there is no need for a worker $w_i$ to report wrong the statistics of its local workload (i.e., the distribution parameters of $\beta_i$) since the asked payment will be large, which may lead to risk of a failure in the offline market. Then, a worker $w_i$ attends the online mode as a temporary worker will not misreport since a large pretended workload can lead to a smaller promised service quality and thus a lower price that a task is willing to pay (due to budget constraint)\footnote{Note that this paper assumes that workers are all truthful. Also, many existing studies \cite{12,43,44} have omit the truthfulness analysis of workers by directly letting them be truthful.}.

\subsection{Service Modeling}
Participating in different tasks can incur various costs for workers, depending on the complexities and resource requirements of each task \cite{12}. Let $c_{i,j,k}$ denote the fundamental cost that $w_i$ needs to spend as long as performing task $\bm{s_{j,k}}$, e.g., energy consumption incurred by traveling around the target sensing region. Also, different workers may offer various service qualities of task processing \cite{12,27,34}, considering factors such as the on-board capabilities of smart devices~\cite{12} (e.g., heterogeneous hardware settings), and channel conditions for sensing/data transmission. In addition, the local workload of each worker impacts its service quality. For example, a worker that assigns top priority to its local workload may cause unacceptable MCSC task completion time. Consequently, we consider two quality levels for the services offered by a worker, namely, \textit{service with hard quality assurance} and \textit{service with soft quality assurance} detailed below.

\noindent
$\bullet$ \textit{\textbf{Service with hard quality assurance}} implies that a worker $w_i\in \mathbb{W}$ offers a strict guaranteed service quality $q_{i,j,k,\mathsf{Hard}}$ to task $\bm{s_{j,k}}$, via promising to assign high priority to the assigned MCSC task under any local workload during each trading ($w_i$ handles its local tasks after the completion of MCSC task). For example, $q_{i,j,k,\mathsf{Hard}}$ can be determined as the theoretically maximum service quality that worker $w_i$ can provide without considering local tasks.
Thus, a worker $w_i$ who offers hard service quality assurance can definitely incur a service cost denoted by $r_iq_{i,j,k,\mathsf{Hard}}\beta_i$, where $r_i$ denotes a positive cost factor, where higher required service quality $q_{i,j,k,\mathsf{Hard}}$ and heavier local workload $\beta_i$ can incur severe cost on worker $w_i$. For example, a higher promised service quality may occupy more resources, while the local tasks should wait for the release of these resources, causing a large cost. From another perspective, the product of the cost factor $r_i$ and the service quality $q_{i,j,k,\mathsf{Hard}}$ can be seen as a unit cost of the occupied resources incurred to the local workload.
Let $p_{i,j,k,\mathsf{Hard}}$ indicate the required payment of worker $w_i$ for contributing data to task $\bm{s_{j,k}}$ under hard quality assurance. 

\noindent
$\bullet$ \textit{\textbf{Service with soft quality assurance}} implies that a worker $w_i$ will assign the  highest priority to its local tasks, while thus offering a fluctuating service quality to MCSC task $\bm{s_{j,k}}$ during each trading (caused by its uncertain local workload). The soft quality is denoted by $q_{i,j,k,\mathsf{Soft}}=q_{i,j,k,\mathsf{Hard}}-r_i^\prime \beta_i$, where, apparently, a large value of workload $\beta_i$ leads to a low value of $q_{i,j,k,\mathsf{Soft}}$. Specifically, $r_i^\prime$ measures the marginal performance degradation rate, capturing the performance degradation of MCSC task caused by local workload (e.g., extra processing time for the assigned MCSC task). Let $p_{i,j,k,\mathsf{Soft}}$ denote the required payment of worker $w_i$ for providing service to task $\bm{s_{j,k}}$ under soft service quality assurance ($p_{i,j,k,\mathsf{Soft}}<p_{i,j,k,\mathsf{Hard}}$). Although soft assurance may incur risky service quality, it can lead to opportunistic advantages for both parties. For instance, the MCSC platform may enjoy a high service quality at a low price (e.g., considering $\beta_i=0$, we have $q_{i,j,k,\mathsf{Soft}}=q_{i,j,k,\mathsf{Hard}}$ and the MCSC platform only pays $p_{i,j,k,\mathsf{Soft}}$). For each $\bm{s_{j,k}}\in\mathbb{S}$, $q_{i,j,k,\mathsf{Soft}}$ also obeys a Truncated normal distribution according to the distribution of $\beta_i\in\mathbb{B}$. 

\section{Proposed Hybrid Worker Recruitment for MCSC}

\subsection{Long-Term Worker Determination and Contract Design (Offline Mode)}
\noindent
We make an rational assumption where the workers and PoIs are independent from each other~\cite{12}, we focus on analyzing the worker recruitment at one PoI and drop the index $j$, without loss of generality
\footnote{Possible collaboration among PoIs and cooperation among workers are not considered in this paper, which will be investigated in our future work as an interesting direction.} 
(e.g., now we have $\mathbb{S}=\{\bm{s_k}|k\in\{1,2,...,|\mathbb{S}|\}\}$, and $\alpha_i\sim\textbf{B}\{(0,1),(1-a_i,a_i)\}$, where “\textbf{B}” denotes the Bernoulli distribution). 


To achieve better analysis, let $\bm{L}=\{\mathsf{Soft},\mathsf{Hard}\}$ denote the set of quality assurance levels, while $\ell\in \bm{L}$ is the corresponding label of ``soft'' and ``hard''. Specifically, $x_{i,k,\mathsf{Hard}}=1$, and $x_{i,k,\mathsf{Soft}}=1$ represent that MCSC platform signs a long-term contract with worker $w_i$ for sensing task $\bm{s_k}$ while offering hard and soft quality assurance, respectively; otherwise $x_{i,k,\mathsf{Hard}}=0$ or $x_{i,k,\mathsf{Soft}}=0$. Also, each worker can be assigned to at most one task under  hard or soft quality assurance ($\sum_{\bm{s_k}\in \mathbb{S}}\sum_{\ell\in\bm{L}}x_{i,k,\ell}\le1, \forall w_i\in \mathbb{W}$); 
while a task can recruit multiple workers. We let $\mathbb{Q}=\{q_{i,k,\ell}|w_i\in \mathbb{W}, \bm{s_k}\in\mathbb{S},\ell\in\bm{L}\}$ indicate the service quality profile, $\mathbb{P}=\{p_{i,k,\ell}|w_i\in\mathbb{W}, \bm{s_k}\in\mathbb{S},\ell\in \bm{L}\}$ represent the payment profile, and $\mathbb{X}=\{x_{i,k,\ell}|w_i\in\mathbb{W}, \bm{s_k}\in\mathbb{S},\ell\in \bm{L}\}$ denote the task assignment profile. Fig. 2 illustrates an example of long-term contract signing conducted at a PoI among four workers and two MCSC tasks. In the following, the utility, expected utility, and risk of participants in the offline trading mode are analyzed. 

\begin{figure}[!t]
\centering
\includegraphics[width=1\linewidth]{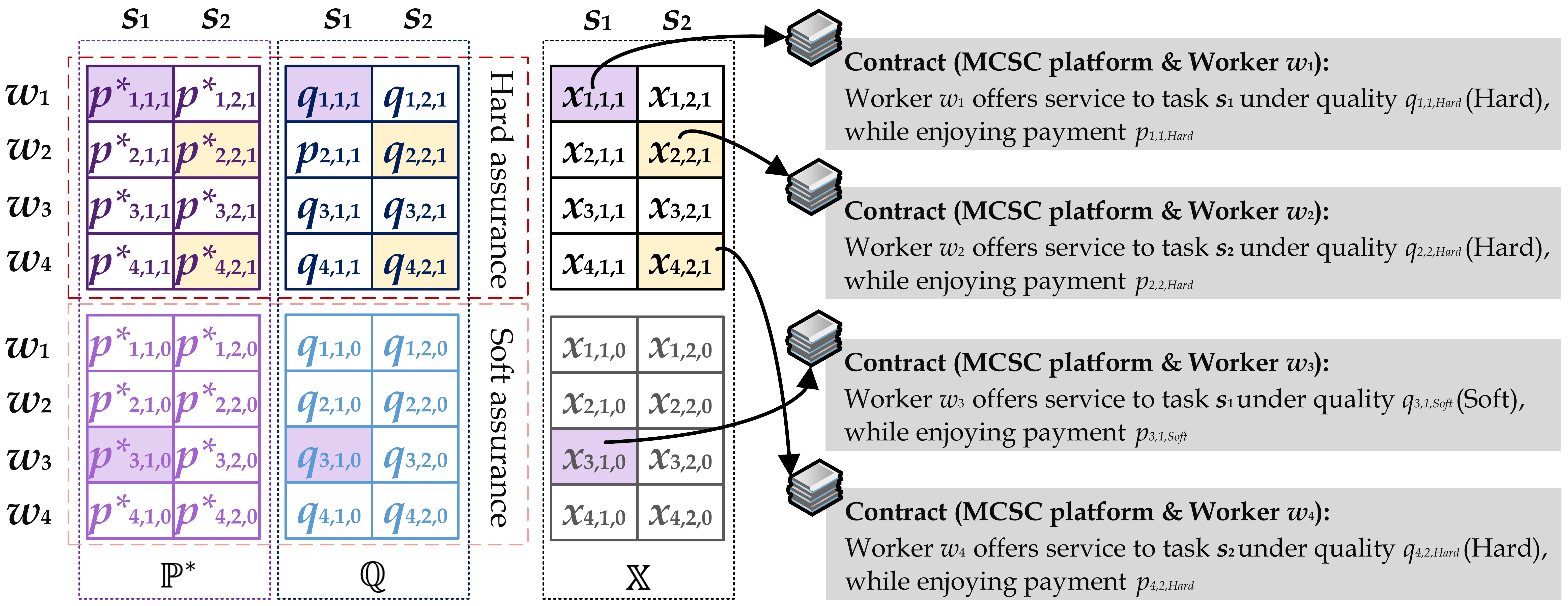}
\caption{Example of long-term contracts among 4 workers and 2 sensing tasks (the third subscript 1, and 0 denote ``Hard'', and ``Soft'', respectively for notational simplicity in this figure).}
\label{fig2}
\end{figure}

\subsubsection{Utility, Expected Utility and Risk}
We define the utility of task $\bm{s_k}\in\mathbb{S}$ as the overall received service quality, as follows, 
\begin{equation}
\label{eq1}
\mathcal{U}^{s_k}\left(\mathbb{X}, \mathbb{Q},\mathbb{A}\right)=\sum_{w_i\in\mathbb{W}}\alpha_i\sum\limits_{\ell\in\bm{L}}x_{i,k,\ell}q_{i,k,\ell},
\end{equation}
where $\alpha_i$ denotes the participation of worker $w_i$, introduced in Sec. 2.3. Correspondingly, the sum utility of all the MCSC tasks is given by
\begin{equation}
\label{eq2}
\mathcal{U}^{\mathbb{S}}\left(\mathbb{X}, \mathbb{Q},\mathbb{A}\right)=\sum_{\bm{s_k}\in \mathbb{S}}\sum_{w_i\in \mathbb{W}}\alpha_i\sum\limits_{\ell\in\bm{L}}x_{i,k,\ell}q_{i,k,\ell}.
\end{equation}
Since it is difficult to directly maximize $\mathcal{U}^{\mathbb{S}}\left(\mathbb{X}, \mathbb{Q},\mathbb{A}\right)$ due to the unpredictable and random nature of MCSC environment (e.g., uncertain $\alpha_i$ and $\beta_i$), we focus on the expected value of $\mathcal{U}^{\mathbb{S}}\left(\mathbb{X}, \mathbb{Q},\mathbb{A}\right)$, assuming independence between $\alpha_i$ and $q_{i,k,\ell}$, $\forall w_i\in\mathbb{W},\bm{s_k}\in \mathbb{S},\ell\in\bm{L}$, obtained as follows: 
\begin{align}
\label{eq3}
&\overline{\mathcal{U}^{\mathbb{S}}}\left(\mathbb{X}, \mathbb{Q},\mathbb{A}\right)=\text{E}\left[\sum_{\bm{s_k}\in \mathbb{S}}\sum_{w_i\in \mathbb{W}}\alpha_i\sum\limits_{\ell\in\bm{L}}x_{i,k,\ell}q_{i,k,\ell}\right] \notag\\
&=\sum_{\bm{s_k}\in \mathbb{S}}\sum_{w_i\in \mathbb{W}}a_i\sum\limits_{\ell\in\bm{L}}x_{i,k,\ell}\overline{q_{i,k,\ell}},
\end{align}
where $\overline{q_{i,k,\ell}}\triangleq\text{E}\left[q_{i,k,\ell}\right]$ denotes the expected value of service quality. For hard quality assurance, we have $\overline{q_{i,k,\mathsf{Hard}}}=q_{i,k,\mathsf{Hard}}$. Also, given the distribution of $\beta_i\in \mathbb{B}$, $q_{i,k,\mathsf{Soft}}$ is a bounded random variable $\mathbbm{q}_{i,k,\mathsf{Soft}}^-\le q_{i,k,\mathsf{Soft}}\le \mathbbm{q}_{i,k,\mathsf{Soft}}^+$, following a normal distribution with mean $\mu_{q_{i,k,\mathsf{Soft}}}=-r_i^\prime \mu_{w_i}+q_{i,k,\mathsf{Hard}}$ and variance ${\left(\sigma_{q_{i,k,\mathsf{Soft}}}\right)}^2={\left(r_i^\prime \sigma_{w_i}\right)}^2$, where $\mathbbm{q}_{i,k,\mathsf{Soft}}^-=-r_i^\prime b_{w_i}^{+}+q_{i,k,\mathsf{Hard}}$ and $\mathbbm{q}_{i,k,\mathsf{Soft}}^+=-r_i^\prime b_{w_i}^{-}+q_{i,k,\mathsf{Hard}}$ are the lower and upper bound of $q_{i,k,\mathsf{Soft}}$, respectively. The expectation of $q_{i,k,\mathsf{Soft}}$ can be calculated as follows
\begin{align}
\label{eq4}
&\overline{q_{i,k,\mathsf{Soft}}}=\mu_{q_{i,k,\mathsf{Soft}}}\notag\\
&+\sigma_{q_{i,k,\mathsf{Soft}}}\frac{\phi\left(\frac{\mathbbm{q}_{i,k,\mathsf{Soft}}^{-}-\mu_{q_{i,k,\mathsf{Soft}}}}{\sigma_{q_{i,k,\mathsf{Soft}}}}\right)-\phi\left(\frac{\mathbbm{q}_{i,k,\mathsf{Soft}}^{+}-\mu_{q_{i,k,\mathsf{Soft}}}}{\sigma_{q_{i,k,\mathsf{Soft}}}}\right)}{\Phi\left(\frac{\mathbbm{q}_{i,k,\mathsf{Soft}}^{+}-\mu_{q_{i,k,\mathsf{Soft}}}}{\sigma_{q_{i,k,\mathsf{Soft}}}}\right)-\Phi\left(\frac{\mathbbm{q}_{i,k,\mathsf{Soft}}^{-}-\mu_{q_{i,k,\mathsf{Soft}}}}{\sigma_{q_{i,k,\mathsf{Soft}}}}\right)},\tag{4}
\end{align}
where $\phi()$ and $\Phi()$ indicate the probability density function (PDF), and cumulative distribution function (CDF) of standard normal distribution. 

Since long-term employment contracts are signed among the MCSC platform and feasible workers in advance to the future practical trading, potential risks should be carefully considered when designing the corresponding contracts. Specifically, for task $\bm{s_k}\in\mathbb{S}$, we consider two major risks as detailed below.

\noindent
$\bullet$\textbf{\textit{Risk of unsatisfying service quality.}} Each task is facing the risk of receiving an unsatisfying service quality, mainly due to possible ``no show'' of long-term workers. We formulate this risk as the probability of obtaining a service quality less than $d_k^{Q}$: 
\begin{align}
\label{eq5}
&\mathcal{R}_1^{s_k}\left(\mathbb{X}, \mathbb{Q},\mathbb{A}\right)=\operatorname{Pr}\left\{\frac{\sum\limits_{w_i\in\mathbb{W}}\sum\limits_{\ell\in\bm{L}}\alpha_ix_{i,k,\ell}q_{i,k,\ell}}{d_k^{Q}}\le \lambda_1^{s_k}\right\},\tag{5}
\end{align}
where $\lambda_1^{s_k}$ denotes a positive threshold coefficient. In (5), a larger value of $\mathcal{R}_1^{s_k}$ indicates a higher risk of receiving unsatisfying quality for task $\bm{s_k}$.

\noindent
$\bullet$\textbf{\textit{ Over-budget risk.}}	The existing uncertainties (e.g., in the value of $\alpha_i$ and $\beta_i$) call for overbooking workers to complete MCSC tasks \cite{12} under desired service qualities. For example, in a trading process, “no show” of the long-term workers may lead to unsuccessful execution of a task. Nevertheless, tasks are generally constrained by their budgets, limiting the number of employed workers. As a result, each task $\bm{s_k}\in\mathbb{S}$ is risking overpaying to workers, caused by overbooking. Thus, we formulate the over-budget risk of $\bm{s_k}$ as the probability of the payment to the employed workers exceeding the pre-determined budget $d_k^{B}$:
\begin{align}
\label{eq6}
&\mathcal{R}_2^{s_k}\left(\mathbb{X}, \mathbb{P},\mathbb{A}\right)=\operatorname{Pr}\left\{\frac{\sum\limits_{w_i\in \mathbb{W}}\sum\limits_{\ell\in\bm{L}}\alpha_ix_{i,k,\ell}p_{i,k,\ell}}{d_k^{B}}> \lambda_2^{s_k}\right\},\tag{6}
\end{align}
where $\lambda_2^{s_k} \ge 1$ represents a positive threshold coefficient.

We then define the utility of worker $w_i\in \mathbb{W}$ during each trading as the net profit for offering sensing and computing service to the assigned MCSC task:
\begin{align}
\label{eq7}
&\mathcal{U}^{w_i}\left(\mathbb{X}, \mathbb{Q},\mathbb{P},\alpha_i,\beta_i\right)\notag\\
&=\alpha_i\sum_{\bm{s_k}\in\mathbb{S}}x_{i,k,\mathsf{Hard}}(p_{i,k,\mathsf{Hard}}-c_{i,k}-\xi_ir_iq_{i,k,\mathsf{Hard}}\beta_i)\notag\\
&+\alpha_i\sum_{\bm{s_k}\in\mathbb{S}}x_{i,k,\mathsf{Soft}}(p_{i,k,\mathsf{Soft}}-c_{i,k})\tag{7},
\end{align}
where $\xi_i$ denotes the weight coefficient, and $r_i q_{i,k,\mathsf{Hard}}\beta_i$ represents the extra local cost incurred by performing task $\bm{s_k}$ with high priority. Maximizing (7) directly is challenging due to the uncertainties, thus, we focus on the expected utility of $w_i$ given by
\begin{align}
\label{eq8}
&\overline{\mathcal{U}^{w_i}}\left(\mathbb{X}, \mathbb{Q},\mathbb{P},\alpha_i,\beta_i\right)\notag\\
&=a_i\sum_{\bm{s_k}\in\mathbb{S}}x_{i,k,\mathsf{Hard}}(p_{i,k,\mathsf{Hard}}-c_{i,k})\notag\\
&-\xi_ia_ir_i\sum_{\bm{s_k}\in\mathbb{S}}x_{i,k,\mathsf{Hard}}q_{i,k,\mathsf{Hard}}\overline{\beta_i}\notag\\
&+a_i\sum_{\bm{s_k}\in\mathbb{S}}(x_{i,k,\mathsf{Soft}}(p_{i,k,\mathsf{Soft}}-c_{i,k})),\tag{8}
\end{align}
where $\overline{\beta_i}\triangleq\text{E}[\beta_i]=\mu_{w_i}+\sigma_{w_i}\frac{\phi\left(\mathbbm{b}^{-}_{w_i}\right)-\phi\left(\mathbbm{b}^{+}_{w_i}\right)}{\Phi\left(\mathbbm{b}^{+}_{w_i}\right)-\Phi\left(\mathbbm{b}^{-}_{w_i}\right)}$, $\mathbbm{b}^{+}_{w_i}=\frac{b^{+}_{w_i}-\mu_{w_i}}{\sigma_{w_i}}$ and $\mathbbm{b}^{-}_{w_i}=\frac{b^{-}_{w_i}-\mu_{w_i}}{\sigma_{w_i}}$. 

On the other hand, each worker $w_i\in \mathbb{W}$ faces the risk of receiving negative utility due to its dynamic local workload. We formulate this risk as the probability of $w_i$’s utility being less than the acceptable minimum value $\mathcal{U}^{min}>0$ when participating in the trading as follows:
\begin{align}
\label{eq9}
\mathcal{R}^{w_i}\left(\mathbb{X}, \mathbb{Q},\mathbb{P},\beta_i\right)=\text{Pr}\left\{\frac{\mathcal{U}^{w_i}\left(\mathbb{X}, \mathbb{Q},\mathbb{P},\alpha_i,\beta_i\right)}{\alpha_i\mathcal{U}^{min}}\le\lambda_1^{w_i}\right\}\tag{9}.
\end{align}


\subsubsection{Problem Design}
\textbf{Problem Formulation.} The long-term worker recruitment problem aims to design feasible long-term contracts, each of which contains: \textit{i)} task-worker execution configuration, (i.e., $x_{i,k,\ell}$, $\forall w_i\in\mathbb{W}, \bm{s_k}\in\mathbb{S}, \ell\in\bm{L}$), \textit{ii)} service quality level (i.e., hard or soft), and \textit{iii)} the corresponding payment $p_{i,k,\ell}$, $\forall w_i\in\mathbb{W}, \bm{s_k}\in\mathbb{S}, \ell\in\bm{L}$). Such a problem is formulated by the following optimization $\bm{\mathcal{P}_0}$ (given in (10)), aiming to maximize the overall expected utility of the MCSC platform (i.e., the expected value of the sum service quality of tasks), while protecting the expectation of profits of workers (constraint (C6)).
\begin{align*}
&\qquad\qquad\bm{\mathcal{P}_0:}
\max\limits_{\mathbb{X}}~\overline{\mathcal{U}^{\mathbb{S}}}\left(\mathbb{X},\mathbb{Q}, \mathbb{A}\right) \tag{10}\\
&\textrm{s.t.}\\
&\text{(C1)}:{\mathcal{R}}_1^{s_k}\left(\mathbb{X}, \mathbb{Q}, \mathbb{A}\right)\le {\lambda}_3^{s_k}, \forall \bm{s_k}\in\mathbb{S}\\
&\text{(C2)}:{\mathcal{R}}_2^{s_k}\left(\mathbb{X}, \mathbb{P}, \mathbb{A}\right)\le {\lambda}_4^{s_k}, \forall \bm{s_k}\in\mathbb{S}\\
&\text{(C3)}:{\mathcal{R}}^{w_i}\left(\mathbb{X}, \mathbb{Q}, \mathbb{P},\beta_i\right)\le {\lambda}_2^{w_i}, \forall w_i\in\mathbb{W}\\
&\text{(C4)}:x_{i,k,\ell}\in \{0,1\}, \forall w_i\in\mathbb{W}, \bm{s_k}\in\mathbb{S}, \ell\in\bm{L}\\
&\text{(C5)}:\sum_{\bm{s_k}\in\mathbb{S}}\sum_{\ell\in\bm{L}}x_{i,k,\ell}\le 1,\forall w_i\in \mathbb{W}\\
&\text{(C6)}: \overline{\mathcal{U}^{w_i}}(\mathbb{X},\mathbb{Q}, \mathbb{P}, \alpha_i, \beta_i)>0, \forall w_i\in\mathbb{W}.
\end{align*}
In $\bm{\mathcal{P}_0}$, $\lambda_3^{s_k}$, $\lambda_4^{s_k}$, and $\lambda_2^{w_i}$ are threshold coefficients within interval $(0,1]$. Also, (C1) determines the acceptable risk of MCSC platform on receiving unsatisfying services; (C2) controls the tolerable soft budget constraint associated with each sensing task; (C3) considers workers' individual rationality, and aims to keep the risk of having unsatisfied workers within a limit. Further, (C4) and (C5) ensure a feasible worker recruitment and task assignment: a worker can at most serve one MCSC task during a trading. More importantly, (C6) ensures that each worker can get non-negative utility from a long-term view. Note that $\bm{\mathcal{P}_0}$ involves solving a complex mixed integer linear programming (MILP) problem, to optimize both continuous (e.g., payment associated with different workers $p_{i,k,\ell}$) and discrete (e.g., task assignment $x_{i,k,\ell}$) variables, which is generally NP-hard. Moreover, our considered risk constraints in probabilistic form (i.e., (C1)-(C3)) can further complicate the problem.

\noindent
\textbf{Problem Transformation.} Since the MCSC platform is generally selfish and prefers to hire workers with lower payments, it can offer a payment which is only slightly higher than what a worker requires (as long as each worker get positive utility, e.g., constraint (C6)), to maximize its utility. Subsequently, constraints (C3) and (C6) can be collectively turned into a fixed payment profile $\mathbb{P}^*=\{p_{i,k,\ell}^*|w_i\in\mathbb{W}, \bm{s_k}\in\mathbb{S},\ell\in\bm{L}\}$, where each element (i.e., $p_{i,k,\mathsf{Hard}}^*$ or $p_{i,k,\mathsf{Soft}}^*$) indicates the acceptable payment of each worker under hard/soft quality assurance (describes why (C6) can be automatically satisfied in $\bm{\mathcal{P}_1}$ by applying $\mathbb{P}^*$, the derivation of which is detailed in Appendix A). Correspondingly, we rewrite problem $\bm{\mathcal{P}_0}$ as $\bm{\mathcal{P}_1}$, where (C2) is transformed into (C7):
\begin{align*}
&\qquad\quad\bm{\mathcal{P}_1:}
\max\limits_{\mathbb{X}}~\overline{\mathcal{U}^{\mathbb{S}}}\left(\mathbb{X},\mathbb{Q}, \mathbb{A}\right)  \tag{11}\\
&\textrm{s.t.}~\text{(C1),~(C4),~(C5)}\\
&\text{(C7)}: \mathcal{R}_2^{s_k}(\mathbb{X},\mathbb{P}^*, \mathbb{A})\le \lambda_4^{s_k}, \forall \bm{s_k}\in\mathbb{S}.
\end{align*}
It is worth noting that transforming $\bm{\mathcal{P}_0}$ to $\bm{\mathcal{P}_1}$ is reasonable in real-life networks since the MCSC platform generally dominates the worker recruitment procedure. Namely, it is always the employer (MCSC platform) who determines which worker to hire, as long as the tolerable payment requirement of each worker can be met (e.g., constraints (C3) and (C6)). 

\noindent
\textbf{Challenges in Solving the Problem.}~Solving $\bm{\mathcal{P}_1}$ is non-trivial since it represents a 0-1 integer linear programming (01ILP) problem with NP-hardness. Furthermore, the constraints of $\bm{\mathcal{P}_1}$ are challenging to satisfy: each of (C1) and (C7) contains $|\mathbb{S}|$ probabilistic inequalities. Besides, elements in $\mathbb{X}$ are inter-inhibitive with each other (e.g., constraint (C5) indicates that if $x_{i,k,\mathsf{Hard}}=1$, then $x_{i,k,\mathsf{Soft}}$ should be 0), which further imposes significant complications during solution design.

To this end, to tackle $\bm{\mathcal{P}_1}$, we first introduce an \underline{e}xhaustive \underline{s}earching \underline{a}lgorithm (ESA) which obtains the optimal solution of $\bm{\mathcal{P}_1}$ by checking all the possible mappings among tasks and workers as well as their service levels. Note that ESA can incur a high computational complexity due to the large search space, it is only effective for small problem sizes (e.g., few tasks and workers). 
Then, to achieve a faster trading response and inspired by the state-of-the-art implicit enumeration method, we further propose a sub-optimal algorithm named \underline{u}nique \underline{i}ndex-based \underline{s}tochastic searching with \underline{r}isk-aware \underline{f}ilter \underline{c}onstraint (UISRFC), in which our analyzed risks are regarded as filters to eliminate unsatisfying solutions while accelerating the convergence. Nevertheless, although UISRFC can handle large size problems (e.g., large number of tasks and workers), its performance is strictly constrained by the number of performed iterations and the randomness of stochastic searching, which offers no optimality guarantee. For example, few iterations may lead to the failure on obtaining a good solution, while a large number of iterations will definitely raise the time spent on decision-making. Therefore, since $\bm{\mathcal{P}_1}$ is highly non-convex, we take one step further and transform it into a more mathematically tractable form to reach better universality in various problem sizes, and develop a \underline{g}eometric \underline{p}rogramming-based \underline{s}uccessive \underline{c}onvex \underline{a}lgorithm (GP-SCA) to solve it. Although it may involve a series of approximations and computations at the platform level, it offers good scalability and works well for diverse problem sizes.


\subsubsection{Design of exhaustive searching algorithm (ESA)}
ESA is an exhaustive search method over the solution space, the psudo-code of which is given in Algo. 1. In Algo. 1, $\bm{\mathcal{X}_n}$ denotes $n$-th solution obtained for $\mathbb{X}$, where $n\in \left\{0,1,2,...,2^{|\mathbb{S}|\times|\mathbb{W}|\times 2}-1\right\}$ indicates the index of solutions. Line 3 shows that each index will be mapped to the corresponding binary number. For example, considering $|\mathbb{S}|=2$, $|\mathbb{W}|=3$, the index $n=5$ can be transferred to a binary number 000000000101, with length $2\times 3\times 2=12$. Although ESA is simple to implement and can reach the optimality of $\bm{\mathcal{P}_1}$, it suffers from high computational complexity of $\mathcal{O}\left(2^{|\mathbb{S}|\times |\mathbb{W}|\times 2}\right)$, which grows exponentially with respect to the value of $|\mathbb{W}|$ and $|\mathbb{S}|$. Also, it will converge when all the possible solutions have been obtained. This makes this algorithm impractical for large problem sizes (e.g., large number of tasks and workers). Motivated by this, we propose UISRFC in the following section. 
\begin{algorithm}[h!t]
\small
\setstretch{0.9} 
\caption{Exhaustive searching algorithm (ESA)}
\SetKwInOut{Input}{Input}\SetKwInOut{Output}{Output}
\Input{$\mathbb{S}$, $\mathbb{W}$, $\mathbb{A}$, $\mathbb{B}$, $\mathbb{Q}$, $\mathbb{P}^*$}
\Output{The optimal solution profile $\mathbb{X}^*$}
Initialization: $\bm{\mathcal{X}_0}\leftarrow\emptyset$, $u_0\leftarrow0$, $n\leftarrow0$,

\For{$1\le n\le 2^{|\mathbb{S}|\times |\mathbb{W}|\times 2}-1$}{

$\bm{\mathcal{X}_n}\leftarrow \text{Decimal-to-Binary}(n)$,~\%~Map index $n$ to the corresponding binary number of length $|\mathbb{S}|\times |\mathbb{W}|\times 2$, and transfer the binary number to the solution set $\bm{\mathcal{X}_n}$

\If {$\bm{\mathcal{X}_n}$ meets constraints (C1), (C5), and  (C7)}{

$u_n\leftarrow\overline{\mathcal{U}^{\mathbb{S}}}\left(\bm{\mathcal{X}_n},\mathbb{Q},\mathbb{A}\right)$

 \If{$u_n<u_{n-1}$}{
  $u_n\leftarrow u_{n-1}$
  
  $\bm{\mathcal{X}_n} \leftarrow \bm{\mathcal{X}_{n-1}} $
 
 }

}
 $n \leftarrow n+1$
}
$\mathbb{X}^*\leftarrow \bm{\mathcal{X}_n}$

\textbf{end algorithm}

\end{algorithm}

\subsubsection{Design of unique index-based stochastic searching with risk-aware filter constraint (UISRFC)}
\noindent
Given the discrete nature of the problem, obtaining feasible solutions for $\bm{\mathcal{P}_1}$ (i.e., $\mathbb{X}$) is one of the most significant difficulties. To this end, inspired by the state-of-the-art implicit enumeration method~\cite{38}, which is a special case of branch-and-bound method, we propose UISRFC algorithm to alleviate the unapplicable computational complexity of ESA, while achieving commendable solutions for $\bm{\mathcal{P}_1}$.

The pseudo-code of UISRFC is given in Algo. 2. UISRFC is an iterative method, where, in each iteration $m$, it first stochastically chooses an index which corresponds to a unique binary number (line 4), and check if it is a feasible solution (lines 5-6). If not, it deletes the index from index set $\bm{N}$ (to achieve unique index property while thus proving time efficiency), and starts another iteration (lines 8-10); if yes, it considers the filter constraints\footnote{A filter constraint generally refers to a constraint that helps filter out bad solutions (e.g., mainly unsatisfied value of the objective function of an optimization problem) before checking other constraints, to accelerate the algorithm.} and checks whether they should be updated (lines 11-18). For example, UISRFC updates the lower bound of filter constraint ${\text{(C}}^f\text{)}$ every time when $\bm{\mathcal{X}_n}$ attains a larger value of (3), as shown in line 13. Based on which, any possible solution that fails to meet ${\text{(C}}^f\text{)}$ will be directly abandoned, without considering other constraints. Furthermore, the upper bound of constraint (C7) should also be adjusted every time a lower over-budget risk has been reached, as given by line 18. Specifically, the update of constraints ${\text{(C}}^f\text{)}$ and (C7) indicates that a better MCSC platforms' utility while a lower risk can be achieved by solution $\bm{\mathcal{X}_n}$, while solution $\bm{\mathcal{X}_{n+1}}$ in the following iteration should be better than  $\bm{\mathcal{X}_n}$; otherwise, it will be dropped directly (lines 9-10). With the above-mentioned operations, UISRFC will finally converge in the direction of larger utility and lower risk. 

Specifically, the computational complexity of UISRFC relies on the number of iterations, as denoted by $\mathcal{O}\left(M\right)$, where $1\le M\le 2^{|\mathbb{S}|\times |\mathbb{W}|\times 2}$. Thus, UISRFC may suffer from performance degradation under a small number of iteration (e.g., it may fail to find a good solution of $\bm{\mathcal{P}_1}$ due to the property of randomness during stochastic searching), upon considering increasing optimization problem sizes. Besides, raising the number of iterations will definitely result in higher running time. For example, considering 5 tasks and 15 workers (namely, $|\mathbb{S}|=5$, $|\mathbb{W}|=15$), there are $2^{5\times 15\times 2}$ potential solutions (although many of them are infeasible) which pose challenges in determining the value of $M$. Also, the operation of stochastic searching of UISRFC brings difficulties in offering optimality guarantee, while its convergence also depends on the pre-determined number of iterations. To design an algorithm with better generality, we further propose GP-SCA to achieve a trackable version of the optimization problem in the following section. 

\begin{algorithm}[t!]
\small
\setstretch{0.9} 
\caption{Unique index-based stochastic searching with risk-aware filtering constraint (UISRFC)}
\SetKwInOut{Input}{Input}\SetKwInOut{Output}{Output}
\Input{$\mathbb{S}$, $\mathbb{W}$, $\mathbb{A}$, $\mathbb{B}$, $\mathbb{Q}$, $\mathbb{P}^*$}
\Output{The optimal solution profile $\mathbb{X}^*$}
Initialization: $\bm{N}\leftarrow\{0,1,2,...,2^{|\mathbb{S}|\times |\mathbb{W}|\times 2}-1\}$; 
$\bm{\mathcal{X}_n}\leftarrow \emptyset, \forall n\in\bm{N}$; 
Iteration $m\leftarrow 1$;
$u^*\leftarrow 0$; $r^*\leftarrow\lambda_4^{s_k}$; Filter constraint associated with utility ${\text{(C}}^f\text{)}: u_n\ge u^*$

\For{$m\le M$}{

\For{$\bm{N}\neq \emptyset$}{
Randomly choose an index $n$ from set $\bm{N}$

$\bm{\mathcal{X}_n}\leftarrow \text{Decimal-to-Binary}(n)$, \%~Map index $n$ to the corresponding binary number of length $|\mathbb{S}|\times |\mathbb{W}|\times 2$; and transfer the binary number to the solution set $\bm{\mathcal{X}_n}$

 \If{$\bm{\mathcal{X}_n}$ meets constraint (C5)}{

$u_n\leftarrow \overline{\mathcal{U}^\mathbb{S}}\left(\bm{\mathcal{X}_n},\mathbb{Q},\mathbb{A}\right)$

\Else{ 
$\bm{N} \leftarrow \bm{N}/\{n\}$

Go to line 20

}

}

\If{${\text{(C}}^f\text{)}$ is satisfied}{
      
       \If{$\bm{\mathcal{X}_n}$ meets constraints (C1), (C7)}{
       $u^*\leftarrow u_n$ \% Update the lower bound of filter constraint ${\text{(C}}^f\text{)}$
       
       $r_n\leftarrow \mathcal{R}^{s_k}_2\left(\bm{\mathcal{X}_n},\mathbb{P}^*,\mathbb{A}\right)$
       
       $\mathbb{X}^*\leftarrow \bm{\mathcal{X}_n}$
       
         \If{$r_n<r^*$}{
         $r^*\leftarrow r_n$
         
         update (C7) by letting $\mathcal{R}^{s_k}_2\left(\mathbb{X},\mathbb{P}^*,\mathbb{A}\right)\le r^*$
         \%~Adjust the upper bound of over-budget risk
         }

      }
}

$\bm{N}\leftarrow \bm{N}/\{n\}$ \%~Avoiding index duplication
}

$m\leftarrow m+1$
}

Output $\mathbb{X}^*$

\textbf{end algorithm}
\end{algorithm}

\subsubsection{Design of geometric programming-based successive convex algorithm (GP-SCA)}
To achieve a trackable optimization problem, we next transform $\bm{\mathcal{P}_1}$ and obtain a tractable solution for it, through a series of convex approximations. Since binary variables in $\mathbb{X}$ pose a great challenge on solving $\bm{\mathcal{P}_1}$, we first propose the following inequality to relax binary variables $x_{i,k,\mathsf{Soft}}$ and $x_{i,k,\mathsf{Hard}}$ to continuous ones within interval $[0,1]$:
\begin{align}
&x_{i,k,\mathsf{Soft}}(1-x_{i,k,\mathsf{Soft}})+x_{i,k,\mathsf{Hard}}(1-x_{i,k,\mathsf{Hard}})\le0, \hspace{-2mm}  \tag{12}
\end{align}
where $x_{i,k,\mathsf{Soft}}$ and $x_{i,k,\mathsf{Hard}}$ can either be 0, or 1 so that to satisfy (12). 
We then reformulate $\bm{\mathcal{P}_1}$ (a maximization problem) to $\bm{\mathcal{P}_2}$ (a minimization problem) given in (13) by introducing variable $y$ and constraint (C9). In $\bm{\mathcal{P}_2}$, constraint (C8) is defined according to (12), while we obtain the tractable versions of (C1) and (C7) based on Markov inequality, given by ${\text{(C1)}}^\prime$ and ${\text{(C7)}}^\prime$, the derivations of which are detailed in Appendix B. 
\begin{align*}
&\qquad\qquad\quad\quad\qquad\qquad \bm{\mathcal{P}_2:} \min\limits_{\mathbb{X}, y}~y\tag{13}\\
&\textrm{s.t.}~\text{(C5)},\\
&{\text{(C1)}}^\prime:(1-\lambda_3^{s_k})\lambda_1^{s_k}d_k^{Q}-{\sum\limits_{w_i\in\mathbb{W}}\sum_{\ell\in\bm{L}}a_ix_{i,k,\ell}\overline{q_{i,k,\ell}}}\le 0,\notag\\&\forall \bm{s_k}\in\mathbb{S}\\
&{\text{(C4)}}^\prime: 0\le x_{i,k,\ell}\le 1, \forall w_i\in\mathbb{W}, \bm{s_k}\in\mathbb{S},l\in\bm{L}\\
&{\text{(C7)}}^\prime:{\sum\limits_{w_i\in\mathbb{W}}\sum_{\ell\in\bm{L}}a_ix_{i,k,\ell}p_{i,k,\ell}^*}-\lambda_2^{s_k}\lambda_4^{s_k}d_k^{B}\le 0,\forall \bm{s_k}\in\mathbb{S}\\
&\text{(C8)}: \sum\limits_{w_i\in\mathbb{W}}\sum\limits_{\bm{s_k}\in\mathbb{S}}\sum_{\ell\in\bm{L}}\left(x_{i,k,\ell}-{(x_{i,k,\ell})}^2\right)\le 0 \\
&\text{(C9)}: \frac{1}{\sum\limits_{\bm{s_k}\in\mathbb{S}}\sum\limits_{w_i\in\mathbb{W}}\sum\limits_{\ell\in\bm{L}} a_i\overline{q_{i,k,\ell}}x_{i,k,\ell} } \le y
\end{align*}
In $\bm{\mathcal{P}_2}$, (C9) constrains the upper bound of $\frac{1}{\overline{\mathcal{U}^{\mathbb{S}}}(\mathbb{X},\mathbb{Q},\mathbb{A})}$ via applying optimization variable $y$, which achieves the same solution with the original maximization problem. In the following, we exploit the innate characteristics of $\bm{\mathcal{P}_2}$ and to transform it to Geometric Programming (GP) format~\cite{39}, the solution of which can be obtained via solving a sequence of convex problems. First, we revisit ${\text{(C1)}}^\prime$, ${\text{(C7)}}^\prime$, (C8), and (C9) and expressing them as normalized inequalities as follows: 
\begin{align}
&{\text{(C1)}}^\prime:\frac{(1-\lambda_3^{s_k})\lambda_1^{s_k}d_k^{Q}}{{\sum\limits_{w_i\in\mathbb{W}}\sum\limits_{\ell\in\bm{L}}a_ix_{i,k,\ell}\overline{q_{i,k,\ell}}}}\le 1,
\forall \bm{s_k}\in\mathbb{S}\tag{14}\\
&{\text{(C7)}}^\prime:\frac{{\sum\limits_{w_i\in\mathbb{W}}\sum\limits_{\ell\in\bm{L}}a_ix_{i,k,\ell}p_{i,k,\ell}^*}}{\lambda_2^{s_k}\lambda_4^{s_k}d_k^{B}}\le 1,\forall \bm{s_k}\in\mathbb{S}\tag{15}\\
&{\text{(C8)}^\prime}:\frac{\sum\limits_{w_i\in\mathbb{W}}\sum\limits_{\bm{s_k}\in\mathbb{S}}\sum\limits_{\ell\in\bm{L}}x_{i,k,\ell}}{\mu+\sum\limits_{w_i\in\mathbb{W}}\sum\limits_{\bm{s_k}\in\mathbb{S}}\sum\limits_{\ell\in\bm{L}}{(x_{i,k,\ell})}^2}\le 1\tag{16}\\
&\text{(C9)}: \frac{1}{y\sum\limits_{\bm{s_k}\in\mathbb{S}}\sum\limits_{w_i\in\mathbb{W}}\sum\limits_{\ell\in\bm{L}} a_i\overline{q_{i,k,\ell}}x_{i,k,\ell} } \le 1,\tag{17}
\end{align}
where $0<\mu\le1$ in (16) represents a constant coefficient approaching to 0, which avoids the tightness of constraint (C8), i.e., the right hand side of (C8) is replaced with $\mu$ instead of 0, which is desired in practical implementation. 

We then focus on the non-convex constraints and aim to approximate them via convex functions. Let functions $f_k^{(C1)}(\bm{x})$, $f^{(C8)}(\bm{x})$, and $f^{(C9)}(\bm{x},y)$ denote the denominator of the fractions in ${\text{(C1)}}^\prime$, $\text{(C8)}^\prime$, and (C9), respectively. We upper bound these functions in (18), (19), and (20) according to arithmetic-geometric mean inequality~\cite{20}, where $\mu^\prime\triangleq\frac{\mu}{|\mathbb{W}|\times|\mathbb{S}|\times |\bm{L}|}$. Specifically, we solve the problem under an iterative manner, where in (18)-(20) we approximate the above functions for variable $\bm{x}=\{x_{i,k,\ell}\}$ around the fixed-point; ${\bm{x}}^{[m]}=\left\{x_{i,k,\ell}^{[m]}\right\}$, where $m$ denotes the index of the iteration, and $x_{i,k,\ell}^{[m]}$ is the solution of the problem at iteration m. Similarly, let $y^{[m]}$ indicate the solution of $y$ at the $m$-th iteration. 
Using these approximations, ${\text{(C1)}}^\prime$, ${\text{(C8)}}^\prime$ and (C9) are further transformed to ${(\hat{\text{C}}\text{1})}^\prime$, ${(\hat{\text{C}}\text{8})}^\prime$, and $(\hat{\text{C}}\text{9})$ given by
\begin{align}
&{(\hat{\text{C}}\text{1})}^\prime:\frac{(1-\lambda_3^{s_k})\lambda_1^{s_k}d_k^{Q}}{\hat{f}^{(C1)}_k(\bm{x})}\le 1,
\forall \bm{s_k}\in\mathbb{S}\tag{21}\\
&{(\hat{\text{C}}\text{8})}^\prime:\frac{\sum\limits_{w_i\in\mathbb{W}}\sum\limits_{\bm{s_k}\in\mathbb{S}}\sum\limits_{\ell\in\bm{L}}x_{i,k,\ell}}{\hat{f}^{(C8)}(\bm{x})}\le 1\tag{22}\\
&{(\hat{\text{C}}\text{9})}:\frac{1}{\hat{f}^{(C9)}(\bm{x},y)}\le 1.\tag{23}
\end{align}
Based on the above steps, $\bm{\mathcal{P}_2}$ can be reformulated a standard GP given by $\bm{\mathcal{P}_3}$:
\begin{align}
&\qquad\qquad\quad\quad\bm{\mathcal{P}_3:} \min\limits_{\mathbb{X},y}~y \tag{24}\\
&\textrm{s.t.}~{(\hat{\text{C}}\text{1})}^\prime, {\text{(C4)}}^\prime, \text{(C5)},  {\text{(C7)}}^\prime, {(\hat{\text{C}}\text{8})}^\prime, (\hat{\text{C}}\text{9})\notag.
\end{align}
Considering the logarithmic change of variables, a standard formed GP can be transformed into a convex optimization problem, which can be solved by using commercial software such as CVX~\cite{20} in an efficient manner. Detailed derivations are shown in Appendix C. 
The corresponding pseudo-code regarding solving $\bm{\mathcal{P}_3}$ is given in Algo. 3. The computational complexity and convergence of GP-SCA relies on the CVX solver and the value of convergence criterion (e.g., when the gap between two adjacent solutions falls below $10^{-4}$). Moreover, by transforming the problem from $\bm{\mathcal{P}_1}$ to $\bm{\mathcal{P}_5}$, some of the original constraints have been loosed which makes the above two optimization problems not strictly the same. Thus, our proposed GP-SCA represents a sub-optimal solution with a tractable form and a better time-efficiency (also proved by simulations).

\subsection{Temporary Worker Determination (Online Mode)}
\noindent
The uncertainties (e.g., $\alpha_i$ and $\beta_i$, $w_i\in\mathbb{W}$) can lead to the case where the actual service quality offered to a task may not reach its satisfaction during each practical trading. For example, the possible absence of some long-term workers associated with task $\bm{s_k}$ during a trading may lead to an unsatisfying overall service quality (e.g., lower than $d_k^{Q}$), while the disbursed expense is within budget $d_k^{B}$. Under this circumstance, the designed online mode can be triggered, where MCSC platform can hire feasible temporary workers, i.e., workers without long-term contracts, for these tasks, via analyzing the present network/market information. Let $\mathbb{S}^\prime$ denote the set of tasks with unsatisfying service quality offered from long-term workers and remaining budgets, and $\mathbb{W}^\prime$ describe the set of workers without signing long-term contracts with the MCSC platform, who have attended the current trading (namely, for all $w_{i^\prime}\in \mathbb{W}^\prime$, we have $\alpha_{i^\prime}=1$). 

The proposed temporary worker determination aims to maximize the overall received quality of tasks in $\mathbb{S}^\prime$, under the current network/market condition, as shown by the following optimization problem.
\begin{align}
&\qquad\quad\bm{\mathcal{P}_4:}
\max\limits_{\mathbb{X}^\prime}~\sum\limits_{\bm{s_{k^\prime}}\in \mathbb{S}^\prime}\sum\limits_{w_{i^\prime}\in \mathbb{W}^\prime}\sum\limits_{\ell\in\bm{L}}x_{i^\prime,k^\prime, \ell}q_{i^\prime,k^\prime, \ell}\tag{25}\\
&\textrm{s.t.}~\notag\\
&\text{(C10)}: x_{i^\prime,k^\prime, \ell}\in \{0,1\}, \forall w_{i^\prime}\in\mathbb{W}^\prime,  \bm{s_{k^\prime}}\in\mathbb{S}^\prime, \ell \in \bm{L}\notag\\
&\text{(C11)}:\sum\limits_{\bm{s_{k^\prime}}\in\mathbb{S}^\prime}\sum\limits_{\ell\in\bm{L}}x_{i^\prime, k^\prime, \ell}\le 1, \forall w_{i^\prime}\in \mathbb{W}^\prime \notag\\
&\text{(C12)}:\sum_{w_{i^\prime}\in\mathbb{W}^\prime}\sum_{\ell\in\bm{L}}x_{i^\prime, k^\prime, \ell}p_{i^\prime, k^\prime, \ell}\le {\left(d^{B}_{k^\prime}\right)}^\prime, \forall \bm{s_{k^\prime}}\in\mathbb{S}^\prime\notag
\end{align}
In $\bm{\mathcal{P}_4}$, $\mathbb{X}^\prime=\{x_{i^\prime, k^\prime, \ell}|w_{i^\prime}\in \mathbb{W}^\prime, \bm{s_{k^\prime}}\in\mathbb{S}^\prime,\ell \in \bm{L}\}$ denotes the profile of temporary worker determination. Specifically, ${\left(d^{B}_{k^\prime}\right)}^\prime, \forall \bm{s_{k^\prime}}\in\mathbb{S}^\prime$ represents the remaining budget, while (C12) indicates the corresponding budget constraint. 
Since $\bm{\mathcal{P}_4}$ represents a 0-1 ILP problem, to follow the similar trading rule in both offline and online modes (and thus achieve the fairness of the market), we design three online algorithms similar to our proposed ESA, UISRFC and GP-SCA (given in Sec. 3.1.3-3.1.5), relying on the current information/data during each practical trading. Due to the similarity, we move the detailed analysis (e.g., the GP format and transferred convex optimization problem associated with $\bm{\mathcal{P}_4}$) to Appendix D due to space limitation. 

\begin{strip}
\hrulefill
\begin{align}
&f_k^{(C1)}(\bm{x})=\sum\limits_{w_i\in\mathbb{W}}\sum\limits_{\ell\in\bm{L}}a_i\overline{q_{i,k,\ell}}x_{i,k,\ell}\Rightarrow f_k^{(C1)}(\bm{x})\ge \hat{f}^{(C1)}_k(\bm{x})\triangleq \prod\limits_{w_i\in\mathbb{W}}\prod\limits_{\ell\in\bm{L}}{\left(\frac{x_{i,k,\ell}f_k^{(C1)}{\left(\bm{x}^{[m-1]}\right)}}{x_{i,k,\ell}^{[m-1]}}\right)}^{\frac{a_i\overline{q_{i,k,\ell}}x_{i,k,\ell}^{[m-1]}}{f_k^{(C1)}\left(\bm{x}^{[m-1]}\right)}}\tag{18}\\
&f^{(C8)}(\bm{x})=\sum\limits_{w_i\in\mathbb{W}}\sum\limits_{\bm{s_k}\in\mathbb{S}}\sum\limits_{\ell\in\bm{L}}\left(\mu^\prime+{(x_{i,k,\ell})}^2\right)
\Rightarrow f^{(C8)}(\bm{x})\ge \hat{f}^{(C8)}(\bm{x})\notag\\&\triangleq \prod\limits_{w_i\in\mathbb{W}}\prod\limits_{\bm{s_k}\in\mathbb{S}}\prod\limits_{\ell\in\bm{L}}{\left(\frac{\left(\mu^\prime+{(x_{i,k,\ell})}^2\right)f^{(C8)}\left(\bm{x}^{[m-1]}\right)}{\mu^\prime+\left({x_{i,k,\ell}^{[m-1]}}\right)^{2}}\right)}^{\frac{\mu^\prime+\left({x_{i,k,\ell}^{[m-1]}}\right)^{2}}{f^{(C8)}\left(\bm{x}^{[m-1]}\right)}}
\tag{19}\\
&f^{(C9)}(\bm{x},y)=\sum\limits_{\bm{s_k}\in\mathbb{S}}\sum\limits_{w_i\in\mathbb{W}}\sum\limits_{\ell\in\bm{L}} ya_i\overline{q_{i,k,\ell}}x_{i,k,\ell}\Rightarrow f^{(C9)}(\bm{x},y) \ge \hat{f}^{(C9)}(\bm{x},y)\notag\\
&\triangleq \prod\limits_{w_i\in\mathbb{W}}\prod\limits_{\bm{s_k}\in\mathbb{S}}\prod\limits_{\ell\in\bm{L}}{\left(\frac{yx_{i,k,\ell}f^{(C9)}\left(\bm{x}^{[m-1]},y^{[m-1]}\right)}{y^{[m-1]}x_{i,k,\ell}^{[m-1]}}\right)}^{\frac{a_i\overline{q_{i,k,\ell}}y^{[m-1]}x_{i,k,\ell}^{[m-1]}}{f^{(C9)}\left(\bm{x}^{[m-1]},y^{[m-1]}\right)}}
\tag{20}
\end{align}
\hrulefill
\end{strip}

\begin{algorithm}[t!]
\small
\setstretch{0.9} 
\caption{Geometric
programming-based successive convex algorithm (GP-SCA)}
\SetKwInOut{Input}{Input}\SetKwInOut{Output}{Output}
\Input{$\mathbb{S}$, $\mathbb{W}$, $\mathbb{A}$, $\mathbb{B}$, $\mathbb{Q}$, $\mathbb{P}^*$, convergence criterion}
\Output{The optimal solution profile $\mathbbm{u}^*$}
Initialization: Iteration count $m=0$

Choose an initial feasible point $\mathbbm{u}^{[0]}$ and $v^{[0]}$

Obtain the approximations given in (18)-(20)

 Replace those approximations in (24)
 
Using the logarithmic change of variables and taking the $\ln$ from constraints, convert the GP programming in (24) to a convex optimization problem (see Appendix C)

$m\leftarrow m+1$

Solve the convex optimization problem using an
arbitrary tool (e.g., CVX~\cite{40}) to obtain the solution $\mathbbm{u}^{[m]}$ and $v^{[m]}$

\If{ the convergence criterion between two consecutive
solutions $\mathbbm{u}^{[m-1]}$ and $\mathbbm{u}^{[m]}$ is not met}{

Go to line 3 and repeat the procedure based on solution $\mathbbm{u}^{[m]}$ and $v^{[m]}$

\Else{
Choose the obtained point as the final solution $\mathbbm{u}^*\leftarrow \mathbbm{u}^{[m]}$

}
}

\textbf{end algorithm}

\end{algorithm}

\begin{figure*}[b!]
\centering
\subfigure[]{\includegraphics[width=.245\linewidth]{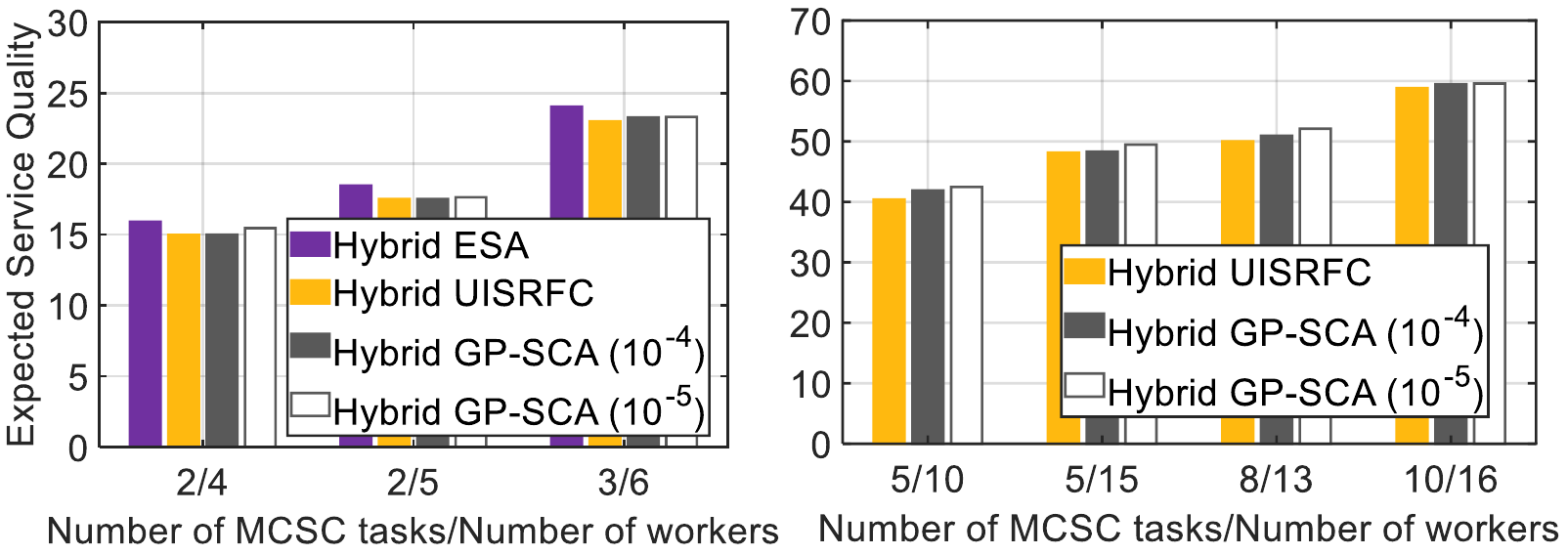}}
\subfigure[]{\includegraphics[width=.245\linewidth,height=.092\linewidth]{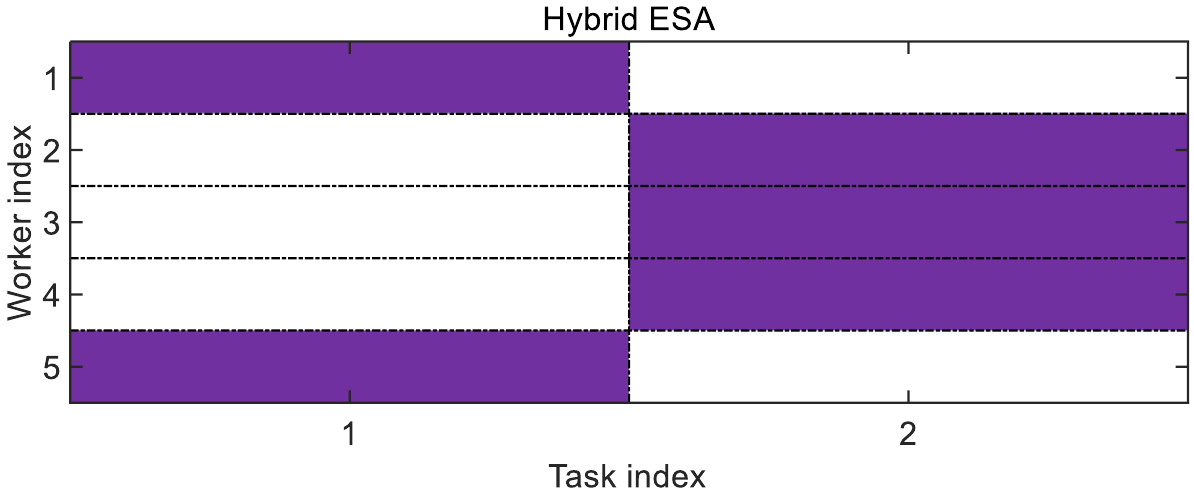}}
\subfigure[]{\includegraphics[width=.245\linewidth,height=.092\linewidth]{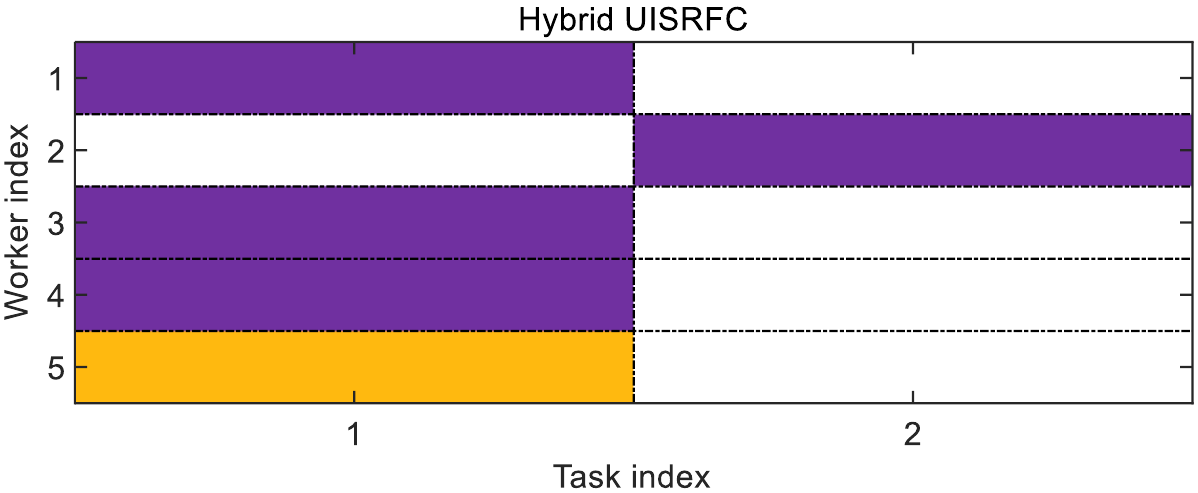}}
\subfigure[]{\includegraphics[width=.245\linewidth,height=.092\linewidth]{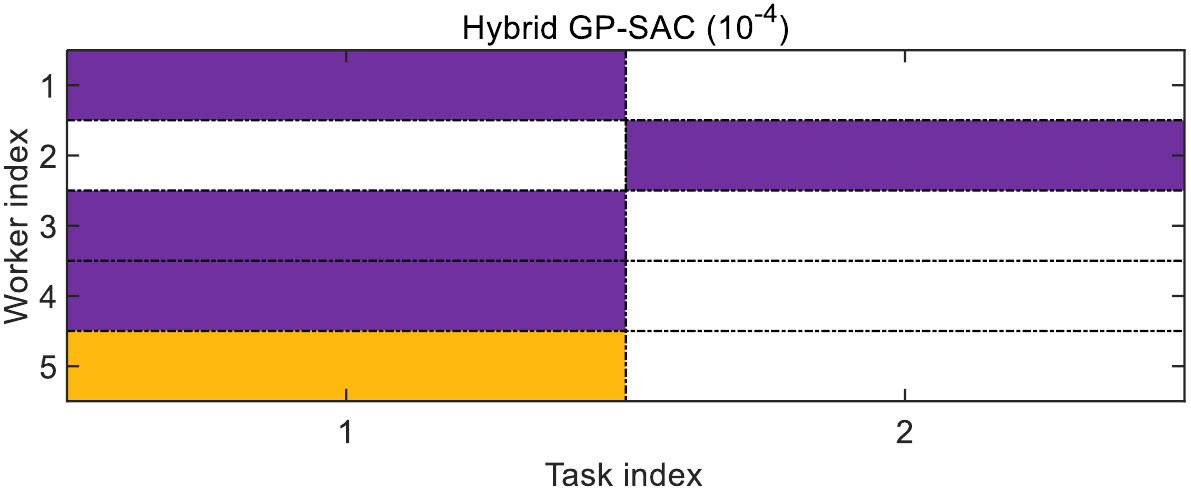}}
\subfigure[]{\includegraphics[width=.245\linewidth,height=.092\linewidth]{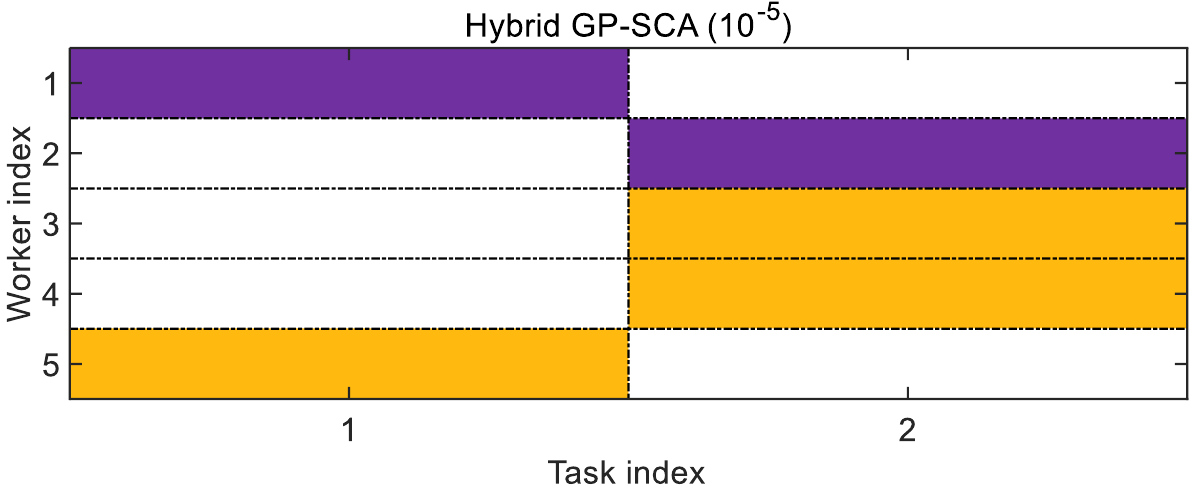}}
\subfigure[]{\includegraphics[width=.245\linewidth,height=.092\linewidth]{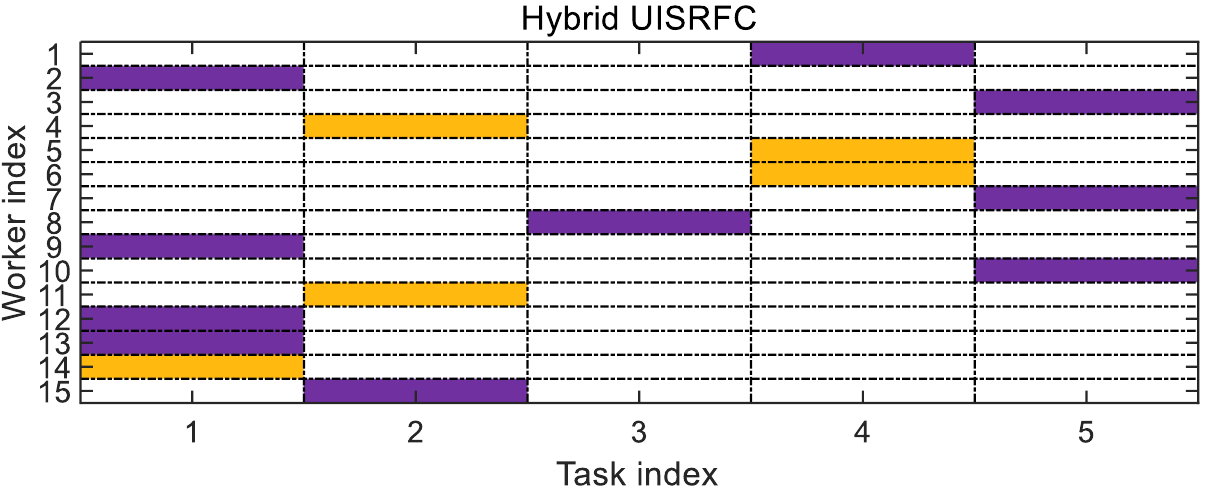}}
\subfigure[]{\includegraphics[width=.245\linewidth,height=.092\linewidth]{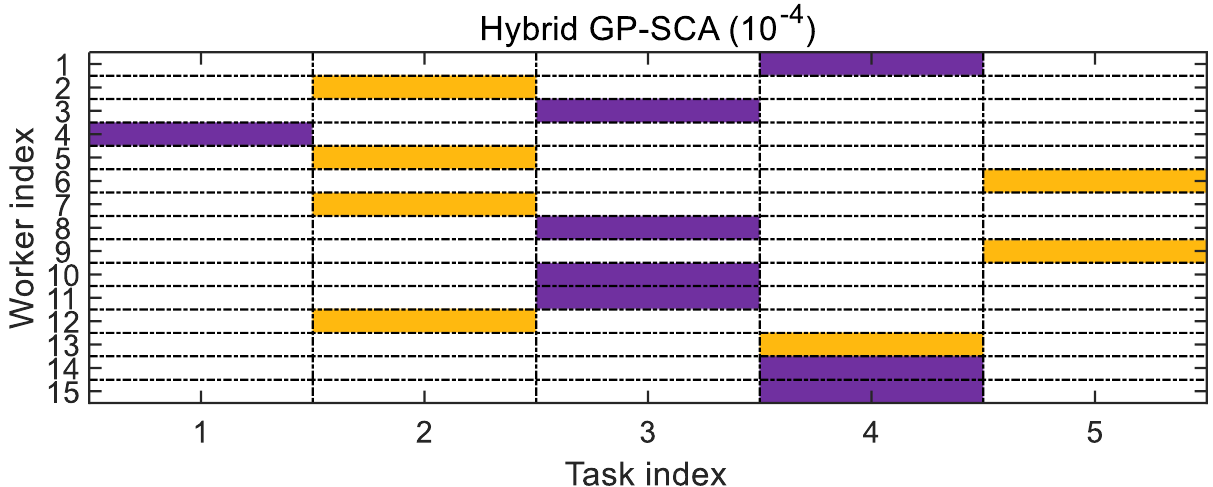}}
\subfigure[]{\includegraphics[width=.245\linewidth,height=.092\linewidth]{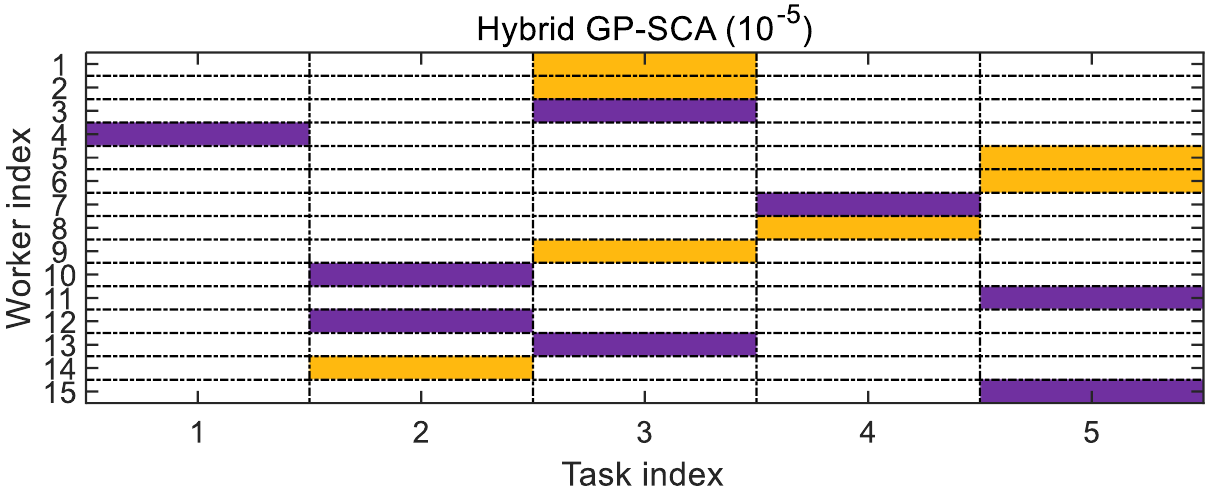}}
\caption{(a). Performance comparison on the expected service quality of MCSC tasks upon considering small, and problem sizes under $\text{Set \#1}$; (b)$\sim$(e) Long-term contract signing associated with 2 tasks and 5 workers in (a); (f)$\sim$(h) Long-term contract signing associated with 5 tasks and 15 workers in (a). Note that figures (b)$\sim$(h), the yellow, and purple-colored box indicates soft, and hard service assurance, respectively.}
\end{figure*}

\vfill

\section{Evaluation}
\noindent
This section presents comprehensive experimental evaluations to demonstrate the validity of our proposed hybrid worker recruitment mechanism for MCSC. To achieve better analysis, we conduct both simulations on numerical data (Section 4.2) and a real-world dataset~\cite{36} (Section 4.3).

\subsection{Benchmark Methods and Basic Parameters}
In our simulations, the proposed algorithms are abbreviated to ``Hybrid ESA'', ``Hybrid UISRFC'', and ``Hybrid GP-SCA'' for notational simplicity. To better verify our commendable performance, 
benchmark methods associated with online trading mode are considered. First, we implement ``Online ESA'', ``Online UISRFC'', and  ``Online GP-SCA'', by drawing inspiration from our proposed solutions. In these modes, all workers are temporary, and each trading is executed based on the current conditions of workers, tasks, and networks. This is achieved through the application of ESA, UISRFC, and GP-SCA, respectively. Then, another two benchmarks derived from the state-of-the-art methods, namely, Monte Carlo random sampling (MCRS)\footnote{Note that the number of iterations of UISRFC and MCRS are set according to different problem sizes, which is lower than or at least equal to $2^{|\mathbb{S}|\times|\mathbb{W}|\times 2}$ (e.g., $\text{min}(10^6, 2^{|\mathbb{S}|\times|\mathbb{W}|\times 2})$.} where task-worker mappings are randomly generated with a certain number of iterations, and service quality-prioritized worker recruitment (SQprefer, inspired by greedy-based algorithms) where each task prefers hard quality assurances, are also considered\footnote{Existing studies, e.g., \cite{38,41,43,47}, with similar ideas on benchmark method design can be supportive.} under an online mode.
 Basic parameters are set as follows: $r_i\in[0.6,0.8]$, $r_i^\prime\in[0.1, 0.3]$, $c_{i,k}\in [1.2, 1.8]$, $q_{i,k,\mathsf{Hard}}\in[3.2,4.8]$, $d_i^{B}\in[15,22]$, $d_i^{Q}\in[7.2, 8.8]$, $\lambda^{s_k}_1\in[1,1.05]$, $\lambda^{s_k}_2\in[0.95,1]$, $\lambda^{s_k}_3\in[0.3,0.4]$, $\lambda^{s_k}_4\in[0.1,0.2]$, $\lambda^{w_i}_1\in[0.98,1.01]$, $\lambda^{w_i}_2\in[0.3,0.4]$.

\subsection{Performance Analysis on Numerical Parameter Settings}
We first conduct simulations on numerical parameters to achieve a better generality of our proposed hybrid service trading mode. To capture diverse behaviors of workers, we consider two sets of parameters to show different distributions of $\alpha_i$ and $\beta_i$, which are $\text{Set \#1}$: $a_i\in\left[64\%,96\%\right]$, $\mu_{w_i}\in [2.4,3.6]$, $\sigma_{w_i}\in[0.4,0.6]$, $b_{w_i}^-\in [\mu_{w_i}-1.44, \mu_{w_i}-0.96]$, $b_{w_i}^+\in [\mu_{w_i}+0.96, \mu_{w_i}+1.44]$; and $\text{Set \#2}$: $a_i\in\left[90\%,95\%\right]$, $\mu_{w_i}\in [2.5,3]$, $\sigma_{w_i}\in[0.1,0.2]$, $b_{w_i}^-\in [\mu_{w_i}-1, \mu_{w_i}-0.5]$, $b_{w_i}^+\in [\mu_{w_i}+0.5, \mu_{w_i}+1]$. Apparently, workers associated with $\text{Set \#1}$ are under a more ``fluctuant'' condition than that of $\text{Set \#2}$ (which are rather ``stable'' and for ``peak demand'' upon having a larger lower bound of $a_i$). By having the above mentioned two sets of distribution parameters (indicating various statistics of uncertainties), the long-term contracts can be update accordingly.

\subsubsection{Expected and practical utility of MCSC platform}
To highlight the benefits of our proposed algorithms and the intriguing hybrid worker recruitment mode for MCSC in wireless networks, we initially present the expected service quality of MCSC tasks during the offline stage, using $\text{Set \#1}$ as an example (see Fig. 3). Note that Fig. 3(a) takes into account both small-size and large-size networks, where online benchmark methods are not considered, as long-term contracts are exclusively associated with the hybrid mode. Besides, Figs. 3(b)-3(h) show the the corresponding contracts, where $10^{-4}$ and $10^{-5}$ in the legend denote the convergence criterion of GP-SCA. For example, the algorithm will stop at the $m$-th iteration when $|y^{[m]}-y^{[m-1]}|\le 10^{-4}$ by setting $10^{-4}$ as the convergence criterion.

As shown in the left-hand plot of Fig. 3(a), which simulates small-size networks (e.g., 2-3 tasks and 4-6 workers), our proposed Hybrid UISRFC and Hybrid GP-SCA can approximate to the performance of ESA, with a relatively low computational complexity (illustrated by Fig. 6). Besides, a tighter convergence criterion (e.g., $10^{-5}$) can help GP-SCA to reach a larger value of expected service quality. Figs. 3(b)-3(e) describe the long-term contracts signing among 2 tasks and 5 workers, where in Hybrid GP-SCA, the expected service quality of MCSC tasks can be improved roughly by $5.5\%$ when applying $10^{-5}$ as the convergence criterion, rather than $10^{-4}$. The right-hand plot of Fig. 3(a) depicts the expected utility of the MCSC platform upon having rather large networks (e.g., 5-10 tasks and 10-16 workers), in which Hybrid ESA has been omitted due to its extremely high computational complexity. Similarly, Hybrid GP-SCA ($10^{-5}$) achieves slightly better performance than GP-SCA with convergence criterion $10^{-4}$, while Hybrid GP-SCA outperforms hybrid URIRFC since the performance of the latter depends heavily on the number of iterations and has randomness due to the large solution space and stochastic searching mode. Figs. 3(f)-3(h) depict how long-term contracts are signed among 5 tasks and 15 workers. Interestingly, our solutions allow a certain overbooking rate to support dynamic service supply since some contractual workers may not be able to offer satisfactory services, due to factors such as their mobility and varying wireless channel qualities. For example, services offered by 4 of the 5 contractual workers $w_2$, $w_9$, $w_{12}$, $w_{13}$, $w_{14}$ associated with task $s_1$ in Fig. 3(f) are sufficient to satisfy $\bm{s_1}$'s service quality requirement, where the overbooking rate is roughly $(5-4)/4=25\%$.

\begin{figure*}[h!t]
\centering
\subfigure[]{\includegraphics[width=0.245\linewidth]{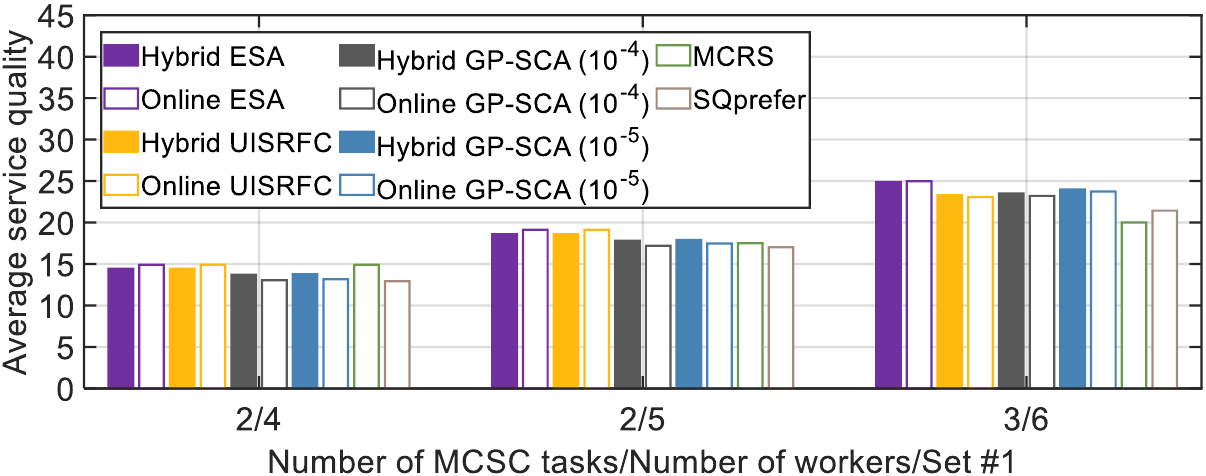}}
\subfigure[]{\includegraphics[width=0.245\linewidth]{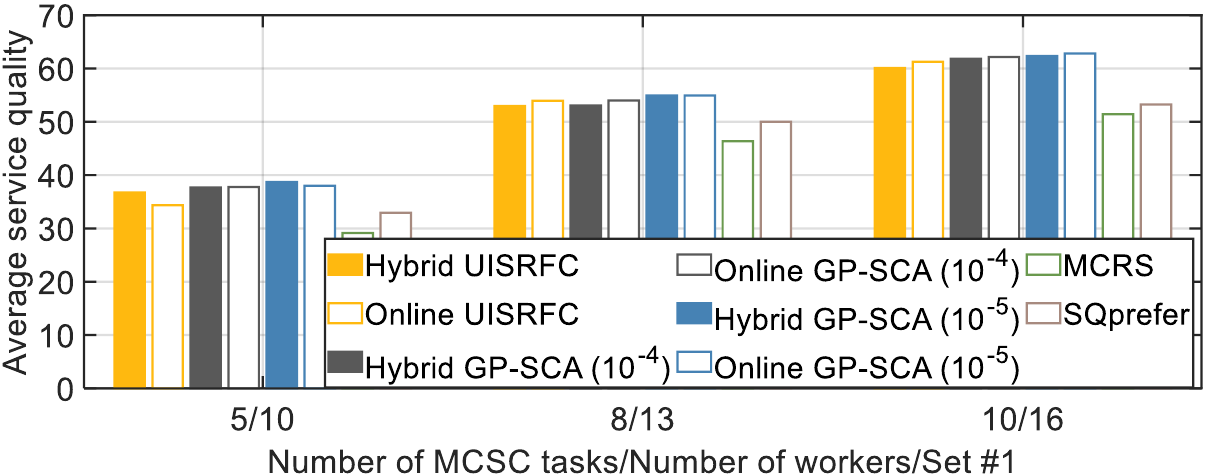}} 
\subfigure[]{\includegraphics[width=0.245\linewidth]{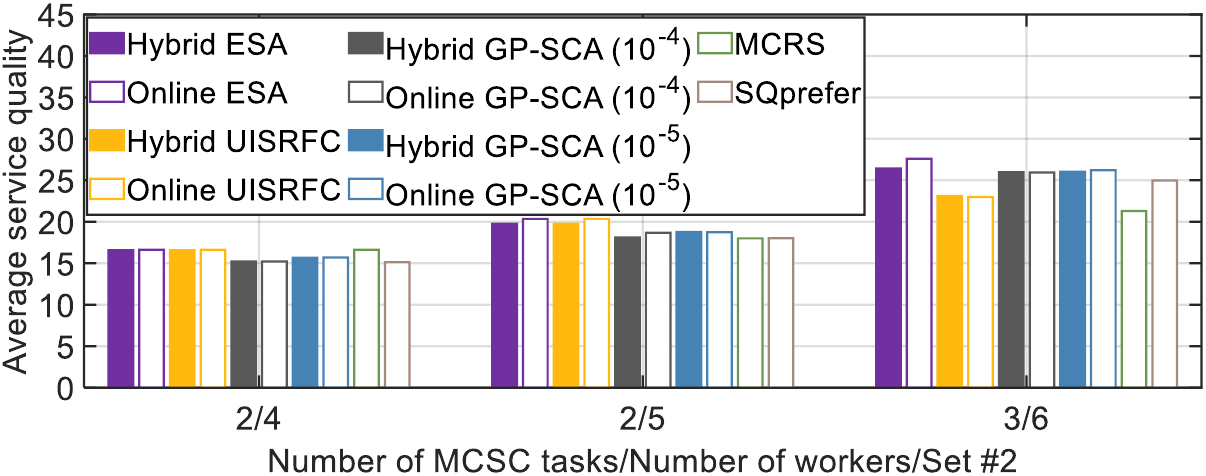}}
\subfigure[]{\includegraphics[width=0.245\linewidth]{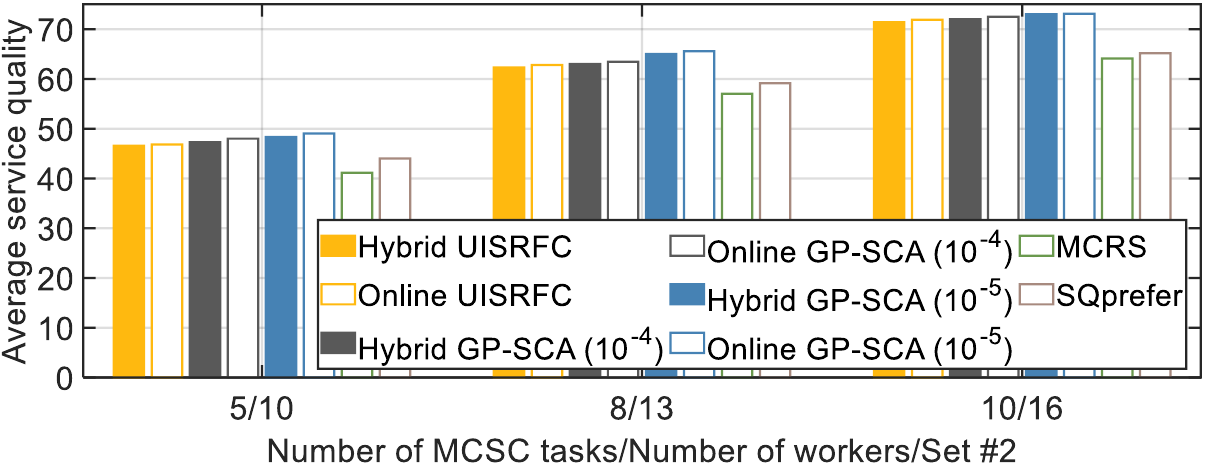}} 
\caption{Performance comparison on the average practical service quality (per trading) of MCSC tasks upon considering different problem sizes, where (a) and (b) rely on $\text{Set \#1}$, while (c) and (d) rely on $\text{Set \#2}$. Note that the average value of each figure is derived from 300 trading. } 
\end{figure*}

\begin{figure*}[h!t]
\centering
\subfigure[]{\includegraphics[width=.245\linewidth]{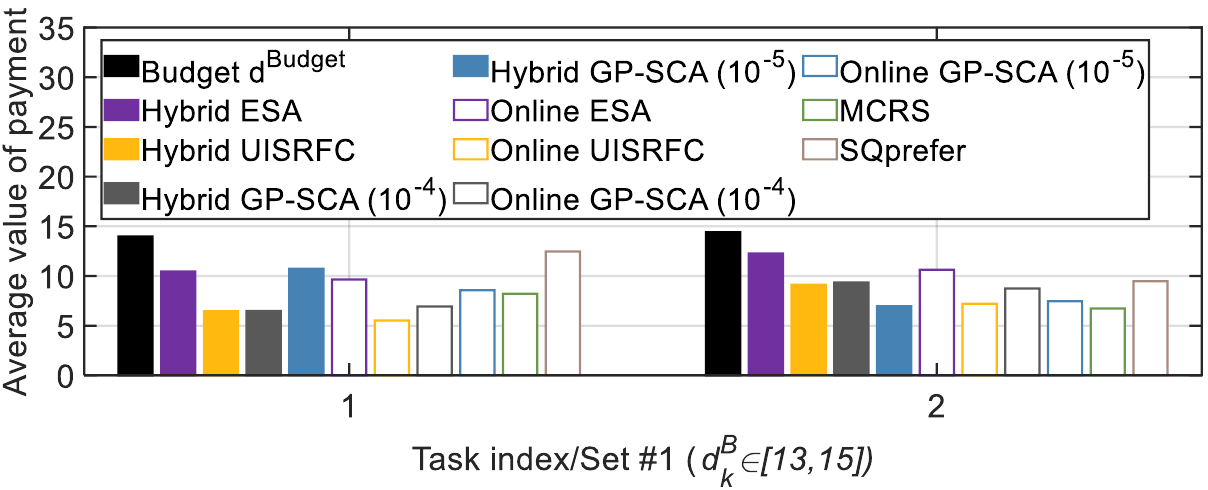}}
\subfigure[]{\includegraphics[width=.245\linewidth]{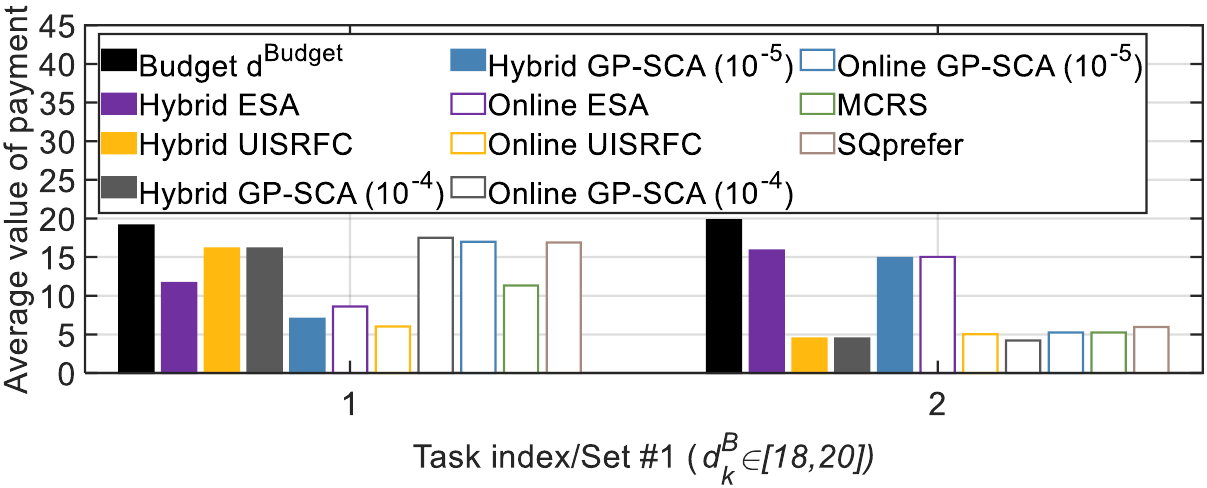}}  
\subfigure[]{\includegraphics[width=.245\linewidth]{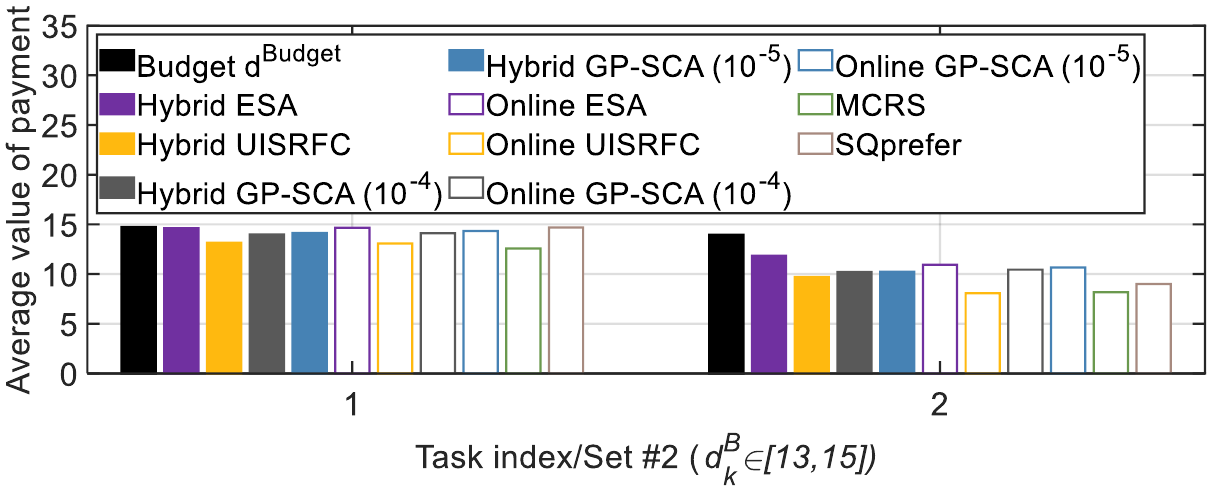}}
\subfigure[]{\includegraphics[width=.245\linewidth]{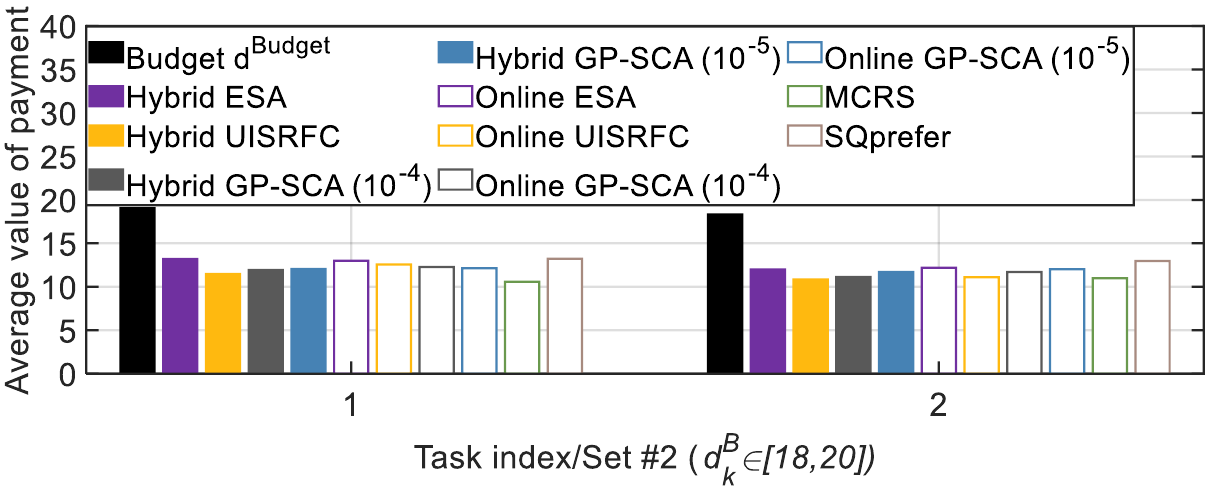}}  
\caption{Performance comparison on $d_k^{B}$ and the average practical expense of MCSC tasks for purchasing workers' services (per trading), upon considering 2 tasks, 6 workers, and different values of budget. Specifically, (a) and (b) rely on $\text{Set \#1}$ upon considering $d^{B}_k\in[13,15]$ and $d^{B}_k\in[18,20]$; while (c) and (d) rely on $\text{Set \#2}$ upon considering $d^{B}_k\in[13,15]$ and $d^{B}_k\in[18,20]$. Note that the average value of each figure is derived from 300 trading.} 
\end{figure*}

We next show the performance on the average practical service quality (per trading) of MCSC tasks from a long-term perspective, in comparison with 5 baseline methods via considering 600 trading in total (i.e., 300 for $\text{Set \#1}$, and 300 for $\text{Set \#2}$, where the long-term contracts have been updated when having a different set). Similarly, different problem sizes are considered to prove the advantages and generalization of our proposed hybrid worker recruitment mechanism. For small problem sizes shown by Figs. 4(a) and Fig. 4(c), although, in most cases, Online ESA performs better due to its optimality under the current network/market situation during each practical trading, our proposed methods reach a far better time efficiency (see Fig. 6). Also, our Hybrid ESA achieves a similar average service quality as compared to Online EAS, since our proposed mechanism allows a certain overbooking rate, thus differentiates which from Online ESA where each trading is implemented under strict budget constraints. Besides, although failing to catch the performance of ESA, our proposed Hybrid UISRFC and Hybrid GP-SCA outperform Online UISRFC and Online GP-SCA (for both $10^{-4}$ and $10^{-5}$ as the convergence criterion) mainly thanks to overbooking. All the considered methods can achieve a larger utility than that of MCSC and SQprefer, due to the randomness and greedy attribute of the two benchmarks.
More importantly, while our proposed methods may exceed the corresponding budget in a specific trading (due to overbooking and probabilistic constraints), the long-term expenses are generally deemed acceptable (refer to Fig. 5), as attributed to the effective management of risks involved.

Fig. 4(b) and Fig. 4(d) illustrate large size problems upon having different sets of distribution parameters (i.e., $\text{Set \#1}$ and $\text{Set \#2}$). In Fig. 4(b), our proposed Hybrid UISRFC and Hybrid GP-SCA can reach a similar average service quality performance to that of the corresponding online trading methods, while sometimes obtaining slightly better performance (e.g., 5 tasks/10 workers, as well as 5 tasks/15 workers in Fig. 4(b)), since the proposed overbooking can help with recruiting more long-term workers, under acceptable over-budget risks. For example, revisiting Fig. 3(h), task $s_5$ can obtain commendable service quality during a trading when all the contractual workers $w_2$, $w_9$, $w_{11}$, $w_{15}$ have participated in trading, which cause an over-budget issue. As for $\text{Set \#2}$ in Fig. 4(d), since workers are more willing to take part in each trading (i.e., $a_i$ is higher than that in $\text{Set \#1}$), the average practical utility of the platform can be higher than that of Fig. 4(b), where the online methods Online UISRFC, Online GP-SCA ($10^{-4}$), and Online GP-SCA ($10^{-5}$) achieve slightly better performance than our proposed hybrid since more feasible solutions can be found when having more available attended workers under budget constraint. However, despite that, our hybrid methods offer a good time efficiency (will be discussed in Fig. 6). Namely, when the service trading market tends to be stable, our hybrid solutions can be closer to the online ones. For example, without uncertainties, e.g., let $a_i=1, \forall w_i\in\mathbb{W}$ and $\beta_i, \forall w_i\in\mathbb{W}$ be a known constant, the online and hybrid methods can almost be the same. This observation further underscores the adaptability of our hybrid methods to dynamic and uncertain market conditions, which are prevalent in real-world network environments. Significantly, the ESA, UISRFC, and GP-SCA methods, implemented in both hybrid and online modes, consistently outperform benchmark methods such as MCRS and SQprefer.

\subsubsection{Budget and payment}
We next test two tasks in Fig. 5, and compares the average practical expense of each MCSC task (per trading) for purchasing sensing and computing services from workers, as well as the corresponding budget, from a long-term perspective over simulating 300 trading for $\text{Set \#1}$, and $\text{Set \#2}$, respectively, upon having different intervals of $d^B_k$. Note that the two tasks tasks are separately shown in Figs. 5(a)-5(d), to achieve a better visuality (i.e., the horizontal coordinate denotes the index of tasks).
As can be seen in Figs. 5(a)-5(b) relying on $\text{Set \#1}$, although overbooking is encouraged in our proposed hybrid worker recruitment mechanism, the average payment of each MCSC task to recruited workers is still acceptable, as benefited from the well-designed risk control constraint (e.g., constraint (C2) of problem $\bm{\mathcal{P}_0}$). For example, considering $\lambda_2^{s_1}=0.98$ and $\lambda_4^{s_1}=0.1$, the risk (namely, probability) of task $\bm{s_1}$'s payment to sensing and computing services exceeds $98\%\times d^{B}_1$ will be controlled within $10\%$. Figs. 5(c)-5(d) under $\text{Set \#2}$ show that the payment from tasks are higher than that of $\text{Set \#1}$ since more workers can attend each practical trading due to a large value of $a_i$. Even so, payments in our proposed hybrid methods will not exceed the corresponding budgets thanks to the well-designed risk management. In addition, our proposed mechanism can achieve a good service quality for tasks while greatly reducing the decision-making time (running time, as illustrated by Fig. 6), below the budgets. 

\begin{figure*}[h!t]
\centering
\subfigure[]{\includegraphics[width=.245\linewidth]{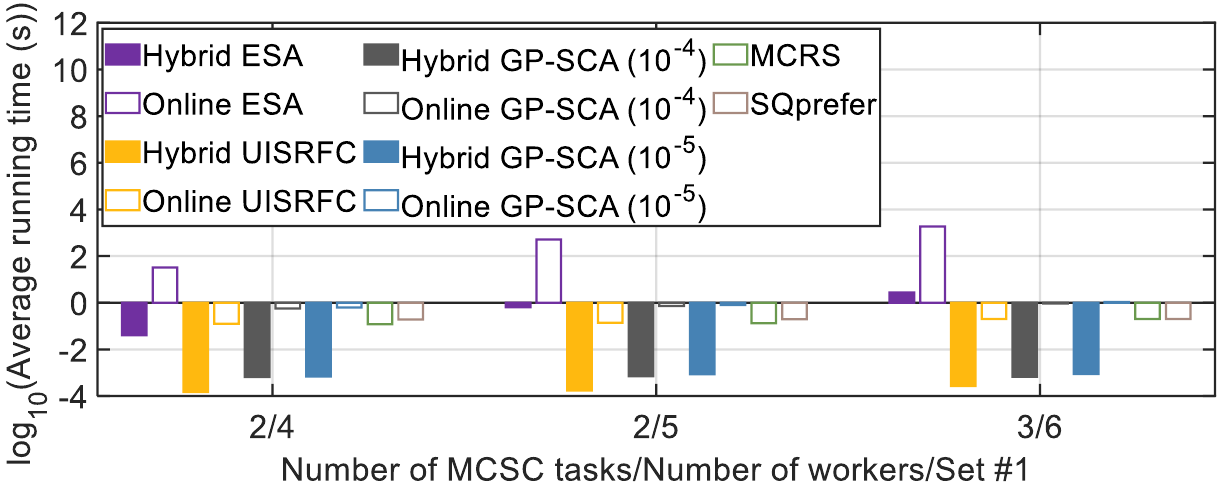}}
\subfigure[]{\includegraphics[width=.245\linewidth]{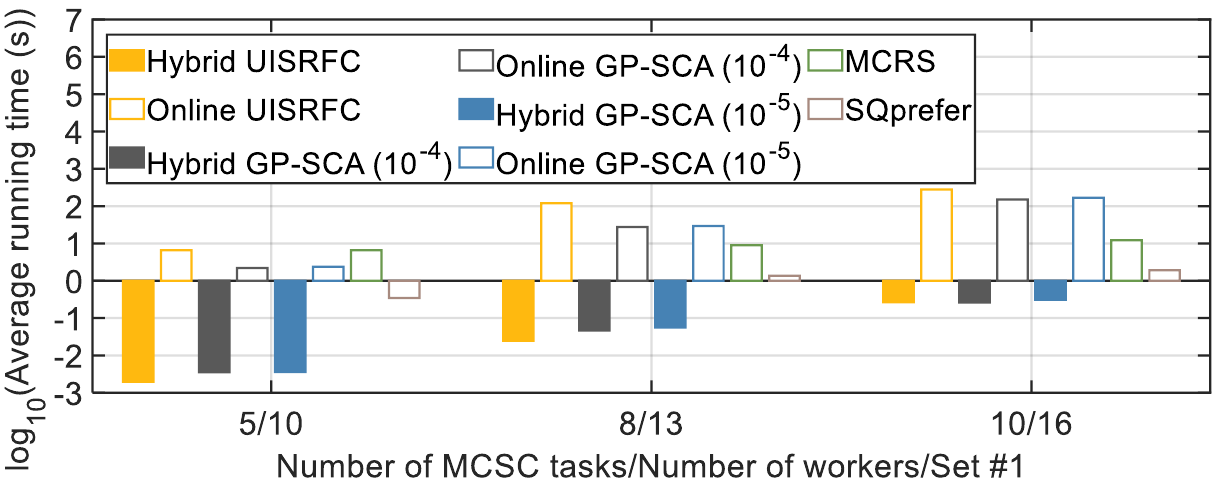}}
\subfigure[]{\includegraphics[width=.245\linewidth]{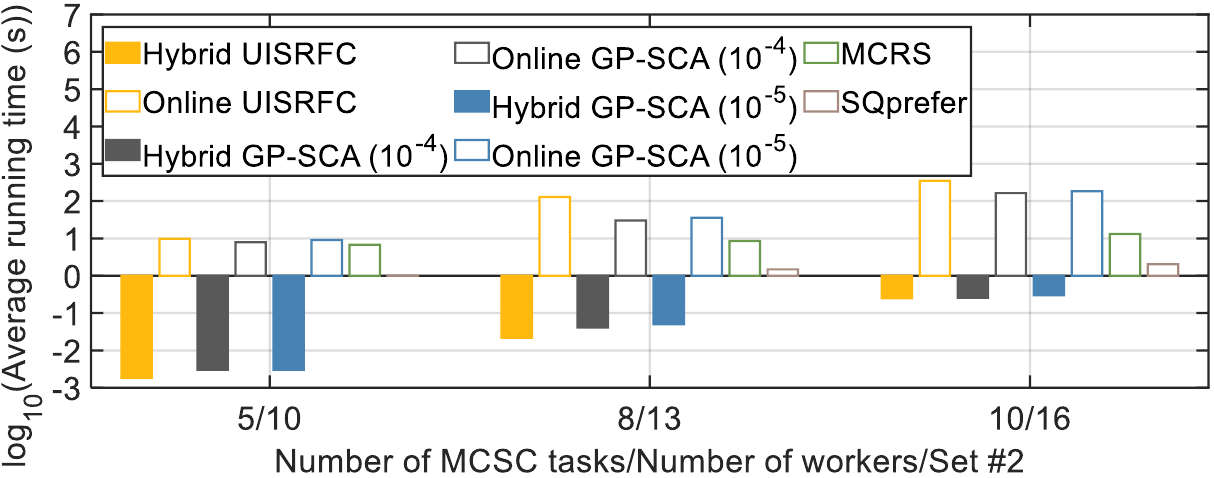}}
\subfigure[]{\includegraphics[width=.245\linewidth]{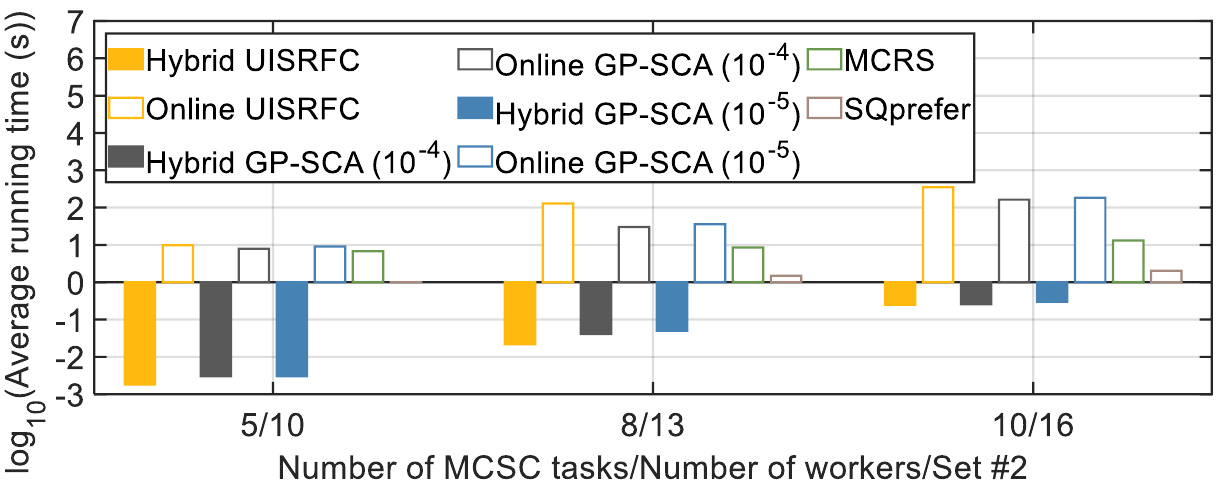}}
\caption{Running time performance upon considering different problem sizes, where (a) and (b) rely on $\text{Set \#1}$, while (c) and (d) rely on $\text{Set \#2}$. Note that the average value of each figure is derived from 300 trading. }
\end{figure*}

\subsubsection{Running time}
Time efficiency represents one of the most important factors in service trading market especially with uncertainties. Thus, we adopt Fig. 6 to analyze the performance on time efficiency measured by the average running time (per trading, which can estimate the time consumed by decision-making), upon having 300 trading for $\text{Set \#1}$ and $\text{Set \#2}$ respectively, while considering small and large problem sizes. Logarithmic function is utilized (in the y-axis of Fig. 6) to better illustrate the gap between different methods. As for small problem sizes in Fig. 6(a) and Fig. 6(c), ESA-based methods always incur a large running time to reach the final solution, where our proposed Hybrid ESA, Hybrid UISRFC and Hybrid GP-SCA greatly outperform the corresponding online methods. 
Moreover, the key reason behind which is that conventional online trading mode should make a decision based on the current network/market information during every trading, which thus incurs a specific delay on decision-making, especially for ESA-based methods. Fortunately, our hybrid methods have determined some long-term worker-task pairs in advance, such an operation can greatly reduce the market scale during each practical trading, as the platform only has to take temporary workers and those tasks with unsatisfying service quality into consideration. 
For example, in Fig. 6(a), the running time associated with Online ESA reaches over 500 seconds averagely during each trading when considering 2 tasks and 5 workers, which makes it impractical to be implemented in real-world networks particularly with energy-constrained mobile devices. In comparison, our Hybrid ESA offers 0.6473 seconds under the same scenario, which thus achieves far better time efficiency and offers a commendable implementation in real-world mobile networks. Although Online UISRFC and Online GP-SCA obtain far better running time performance than Online ESA, the accumulated delay with the increasing number of trading can still bring challenges to the long-term development of MCSC networks. 
For instance, Online GP-SCA ($10^{-5}$) (the scenario of 2 tasks and 4 workers in Fig. 6(a)) causes around 0.5636 seconds decision-making delay on average, that may be practicable during one trading, which, however, will cost around 563.6 seconds over 100 trading. For large problem sizes shown by Fig. 6(b) and Fig. 6(d), our proposed Hybrid UISRFC and Hybrid GP-SCA can still offer commendable running time, as benefited by pre-determined long-term workers. On the contrary, the running time associated with Online UISRFC and Online GP-SCA gradually become prohibitive with increasing number of workers and tasks. Specifically, applying $\text{Set \#2}$ may suffer from an increasing running time rather than $\text{Set \#1}$ e.g., scenario of 10 tasks/16 workers in Fig. 6(b) and Fig. 6(d), since more workers will participant in each practical trading, which thus raise the market scale that the platform needs to handle, especially for online methods. Moreover, although MCRS and SQprefer offer a good performance in time efficiency, they receive unsatisfying practical platform utilities (shown by Fig. 4).

\begin{figure}[h!t]
\centering
\subfigure[]{\includegraphics[width=.49\linewidth]{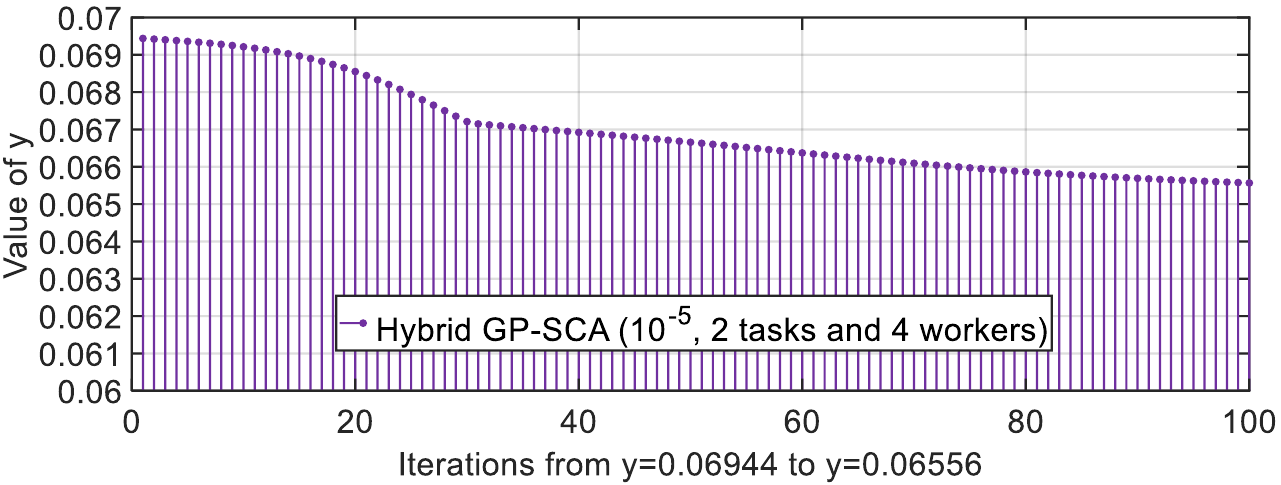}}
\subfigure[]{\includegraphics[width=.49\linewidth]{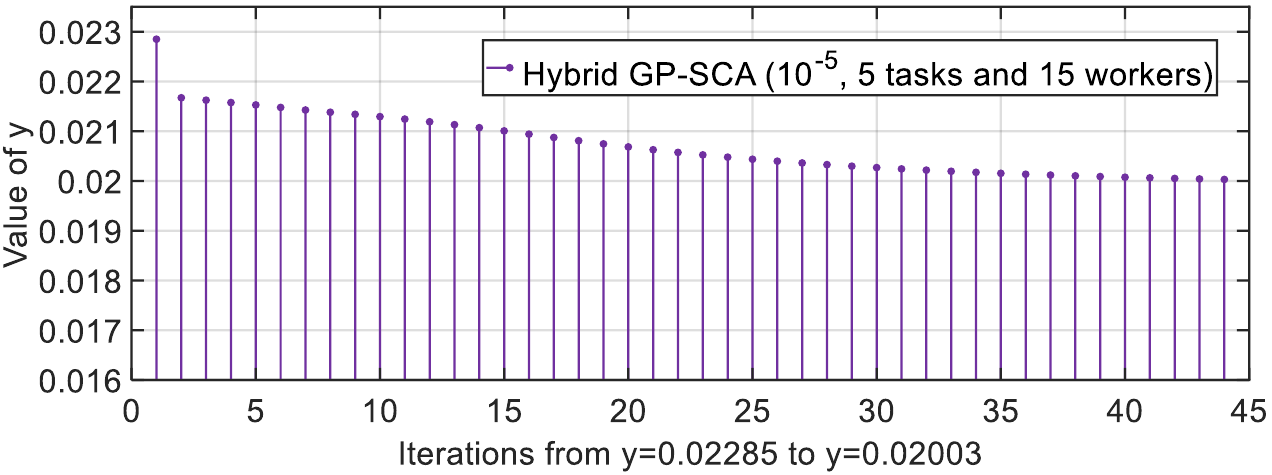}}
\caption{Convergence performance on solving $\bm{\mathcal{P}_3}$ by the proposed GP-SCA regarding $\text{Set \#1}$ and $10^{-5}$ as the convergence criterion.}
\end{figure}

\subsubsection{Convergence of GP-SCA}
Figure 7 illustrates how our GP-SCA-based solution for $\bm{\mathcal{P}_3}$ converges to its given convergence criterion ($10^{-5}$). Note that finding the initial feasible point (e.g., $\mathbbm{u}^{[0]}$ and $v^{[0]}$) is also critical for the convergence speed, for which we show Fig. 7(a) with a rather “bad” initial point, and Fig. 7(b) with a “good” one. As can be seen from these figures, our proposed GP-SCA can exhibit a good performance on convergence, upon considering different problem sizes. 

As a summary, the proposed hybrid worker recruitment methodology for mobile crowd sensing and computing that integrates both online and offline trading modes achieves good performance on service quality in comparison with online trading modes, while offering far better time efficiency, which provides a commendable reference for the continuable development of future MCSC networks. 

\subsection{Performance Analysis on Real-World Dataset}
To better verify the performance of our proposed Hybrid service trading methodology, we then adopt a real-world dataset of Chicago taxi trips~\cite{36}, which records taxi rides in Chicago from 2013 to 2016 with 77 community areas. Here, we consider the $77^\text{th}$ community area as our interested sensing region (e.g., one PoI) and randomly choose 15 taxis as workers, as well as 5 tasks in our simulation (supposed that tasks are randomly distributed in the considered region).
Specifically, we estimate the distribution parameters of $\alpha_i$ by counting the number of days that these taxis arrive at the considered region in January 2013. To better quantize service cost $c_{i,k}$, we record three key factors from the dataset: \textit{i)} the traveled distance of each taxi (i.e., the distance between pick-up locations and drop-off locations); \textit{ii)} the distance between the current positions of taxis (e.g., pick-up location of taxis) and that of each task; \textit{iii)} the distance between the position of a taxi after task completion (e.g., the drop-off location of taxis) and location of the corresponding task. As each $c_{i,k}$ stands for a monetary expression, we make it to be proportional to these above-mentioned factors. Besides, note that since this paper offers some distinct elements, e.g., different levels of services and local workload of workers, not all the data (e.g., the values of soft and hard service quality) can be found in a real-world dataset, we should also consider some manual numerical parameter settings similar to Section 4.2 (e.g., studies \cite{12,46} can be supportive with similar ideas on using both real-world and numerical simulation parameters in MCS networks, especially regarding economic behaviors). For example, the hard service quality is chosen to be inversely proportional to the sum of distances in the previous \textit{ii)} and \textit{iii)}, namely, a long distance leads to a lower quality. Accordingly, we show Fig. 8 to demonstrate our superior performance in comparison with benchmarks. Apparently, in Fig. 8, our proposed Hybrid methods can achieve commendable performance on service quality while maintaining satisfying time efficiency.

\begin{figure}[h!t]
\centering
\subfigure[]{\includegraphics[width=.48\linewidth]{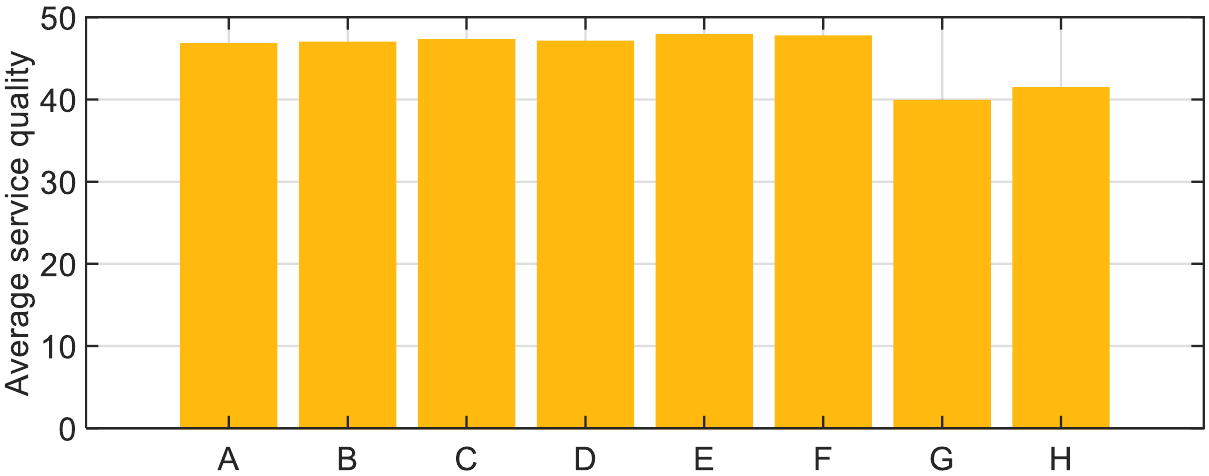}}
\subfigure[]{\includegraphics[width=.49\linewidth]{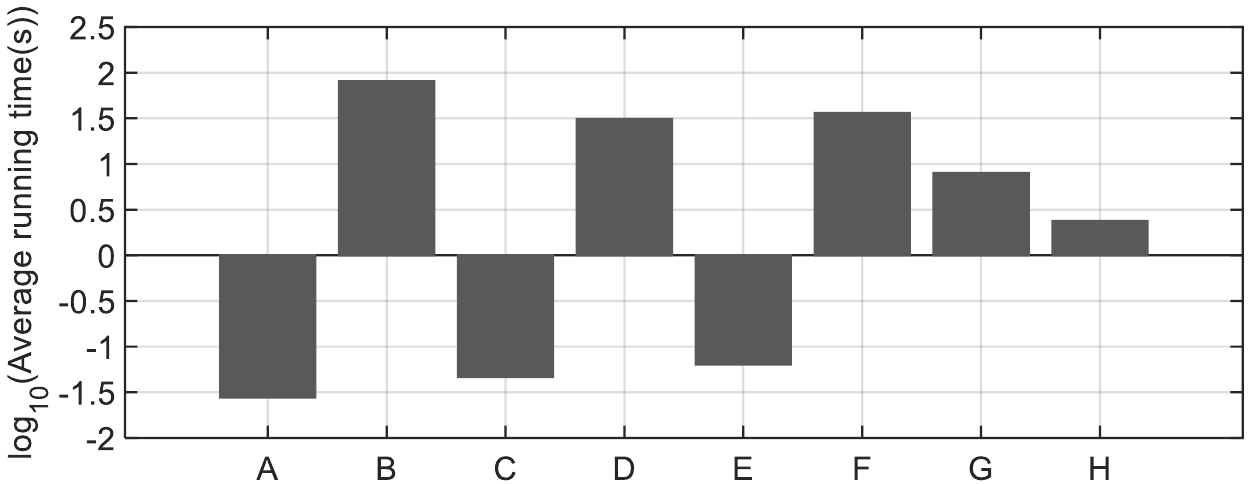}}
\caption{Performance on average service quality and running time  upon having 5 tasks and 15 workers, where A: Hybrid UISRFC, B: Online UISRFC, C: Hybrid GP-SCA ($10^{-4}$), D: Online GP-SCA ($10^{-4}$), E: Hybrid GP-SCA ($10^{-5}$), F: Online GP-SCA ($10^{-5}$), G: MCRS, H: SQprefer}
\end{figure}

\section{Conclusion}
This paper introduces a novel risk-aware hybrid worker recruitment mechanism for MCSC networks, involving diverse uncertainties while integrating both online (long-term worker recruitment) and offline (temporary worker recruitment) trading modes. The optimization targets are formulated as 0-1 integer linear programming problems with NP-hardness. Accordingly, three algorithms, namely, exhaustive searching, unique index-based stochastic searching with risk-aware filter constraint, and geometric programming-based successive convex algorithm, are designed to achieve the corresponding optimal (with high computational complexity) or sub-optimal (with low computation complexity) solutions. Comprehensive simulations demonstrate the effectiveness of our proposed mechanism, in comparison to conventional trading methods. Several promising avenues for future research include exploring worker cooperation, as well as the design of cost-effective and time-efficient algorithms for the challenging NP-hard problems identified. Furthermore, the concept of network environment-aware renewable contract design is currently viewed as an area closely related to our work.

\vfill

 \clearpage
 \newpage
\appendices
\section{Analysis of $\mathbb{P}^*$}
\noindent
For a worker $w_i\in\mathbb{W}$, the tolerable payment $p_{i,k,\mathsf{Hard}}$ and $p_{i,k,\mathsf{Soft}}$ under hard/soft assurance can be determined by constraints (C3) and (C6). 

\noindent
$\bullet$\textbf{Derivation on $p_{i,k,\mathsf{Soft}}^*$ (payment associated with soft assurance):} We first analyze the acceptable payment $p_{i,k,\mathsf{Soft}}^*$ when a worker $w_i$ offers service to sensing task $\bm{s_k}$ under soft quality assurance. Assume that $x_{i,k,\mathsf{Soft}}=1$, risk of worker $w_i$ is calculated by the following (26).
\begin{align}
 \label{eq26}
&\mathcal{R}^{w_i}(x_{i,k,\mathsf{Soft}}=1,q_{i,k,\mathsf{Soft}},p_{i,k,\mathsf{Soft}},\beta_i)\notag\\
&=\operatorname{Pr}\left\{p_{i,k,\mathsf{Soft}}\le \lambda_1^{w_i}\mathcal{U}^{min}+c_{i,k} \right\}\tag{26}
\end{align}
Apparently, $p_{i,k,\mathsf{Soft}}^*$ should satisfy the following inequality (27) to meet constraint (C3),
\begin{align}
 \label{eq27}
p_{i,k,\mathsf{Soft}}^*>\lambda_1^{w_i}\mathcal{U}^{min}+c_{i,k}\tag{27}
\end{align}
In this case, we have the expected utility of worker $w_i$ as given in (28).
\begin{align}
 \label{eq28}
&\overline{\mathcal{U}^{w_i}}(x_{i,k,\mathsf{Soft}}=1,q_{i,k,\mathsf{Soft}},p_{i,k,\mathsf{Soft}},\alpha_i,\beta_i)\notag\\
&=a_i\left(p_{i,k,\mathsf{Soft}}-c_{i,k}\right)\tag{28}
\end{align}
Thus, the acceptable payment $p_{i,k,\mathsf{Soft}}^*$ can be defined by (29), to meeting constraints (C3) and (C6), where $\Delta p$ denotes a standard incentive agreed by all the participants in the MCSC network.
\begin{align}
 \label{eq29}
&p_{i,k,\mathsf{Soft}}^*=\text{max}\left(\lambda_1^{w_i}\mathcal{U}^{min}+c_{i,k}, c_{i,k}\right)+\Delta p\notag\\
&=\lambda_1^{w_i}\mathcal{U}^{min}+c_{i,k}+\Delta p\tag{29}
\end{align}

\noindent
$\bullet$\textbf{Derivation on $p_{i,k,\mathsf{Hard}}^*$ (payment associated with hard assurance):} Assume that $x_{i,k,\mathsf{Hard}}=1$, we have the following (30): 
\begin{align}
 \label{eq30}
&\mathcal{R}^{w_i}\left(x_{i,k,\mathsf{Hard}}=1, q_{i,k,\mathsf{Hard}}, p_{i,k,\mathsf{Hard}}, \alpha_i, \beta_i\right)\notag\\
&=\operatorname{Pr}\left\{-\xi_ir_iq_{i,k,\mathsf{Hard}}\beta_i+p_{i,k,\mathsf{Hard}}-c_{i,k}\le \lambda_i^{w_i}\mathcal{U}^{min}\right\}\notag\\
&=1-\operatorname{Pr}\left\{\xi_ir_iq_{i,k,\mathsf{Hard}}\beta_i\le p_{i,k,\mathsf{Hard}}c_{i,k}-\lambda_i^{w_i}\mathcal{U}^{min}\right\}\tag{30}
\end{align}

Let random variable $Y=\xi_i q_{i,k,\mathsf{Hard}} r_i \beta_i$ for notational simplicity. According to the distribution of $\beta_i$, we have $Y$ conditional on $\mathbbm{y}^{-}\le Y\le \mathbbm{y}^{+}$ associated with Normal Distribution with mean $\mu_Y=\xi_i r_i q_{i,k,\mathsf{Hard}} \mu_{w_i}$ and variance ${(\sigma_Y )}^2={(\xi_i r_i q_{i,k,\mathsf{Hard}} \sigma_{w_i})}^2$, which thus follows a Truncated Normal Distribution denoted by $Y\sim\text{N}\left(\mu_Y,{(\sigma_Y)}^2,\mathbbm{y}^{-},\mathbbm{y}^{+} \right)$. Specifically, $\mathbbm{y}^{-}=\xi_i r_i q_{i,k,\mathsf{Hard}} b_{w_i}^{-}$, and $\mathbbm{y}^{+}=\xi_i r_i q_{i,k,\mathsf{Hard}} b_{w_i}^{+}$ indicates the lower and upper bound of Y, respectively. Thus, CDF of $Y$ is given in (31).
\begin{align}
 \label{eq31}
\text{F}_Y(Y\le y)=\begin{cases}
0,~~~y<\mathbbm{y}^{-}\\
\frac{\Phi\left(\frac{y-\mu_Y}{\sigma_Y}\right)-\Phi\left(\frac{\mathbbm{y}^{-}-\mu_Y}{\sigma_Y}\right)}{\Phi\left(\frac{\mathbbm{y}^{+}-\mu_Y}{\sigma_Y}\right)-\Phi\left(\frac{\mathbbm{y}^{-}-\mu_Y}{\sigma_Y}\right)}\\
1,~~~y>\mathbbm{y}^{+}
\end{cases}\tag{31}
\end{align}
According to (31), (30) can be rewritten by (32),
\begin{align}
 \label{eq32}
&\mathcal{R}^{w_i}\left(x_{i,k,\mathsf{Hard}}=1, q_{i,k,\mathsf{Hard}}, p_{i,k,\mathsf{Hard}}, \alpha_i, \beta_i\right)\notag\\
&=1-\text{F}_Y\left(Y\le p_{i,k,\mathsf{Hard}}-c_{i,k}-\lambda_1^{w_i}\mathcal{U}^{min}\right)\notag\\
&\begin{cases}
0, p_{i,k,\mathsf{Hard}}-c_{i,k}-\lambda_1^{w_i}\mathcal{U}^{min}>\mathbbm{y}^{+}  \\
1-\frac{\Phi\left(\frac{p_{i,k,\mathsf{Hard}}-c_{i,k}-\lambda_1^{w_i} \mathcal{U}^{min}-\mu_Y}{\sigma_Y}\right)-\Phi\left(\frac{\mathbbm{y}^{-}-\mu_Y}{\sigma_Y}\right)}{\Phi\left(\frac{\mathbbm{y}^{+}-\mu_Y}{\sigma_Y}\right)-\Phi\left(\frac{\mathbbm{y}^{-}-\mu_Y}{\sigma_Y}\right)}, \\
 \mathbbm{y}^{-}\le p_{i,k,\mathsf{Hard}}-c_{i,k}-\lambda_1^{w_i}\mathcal{U}^{min}\le \mathbbm{y}^+\\
1, p_{i,k,\mathsf{Hard}}-c_{i,k}-\lambda_1^{w_i}\mathcal{U}^{min}<\mathbbm{y}^{-} 
\end{cases}\tag{32}
\end{align}

In constraint (C3), $\lambda_2^{w_i}\in[0,1)$ where $\lambda_2^{w_i}=0$ means worker $w_i$ does not accept any risk. Consequently, $p_{i,k,\mathsf{Hard}}^*$ should meet the following inequalities:
\begin{equation}
\label{eq33}
\left\{\begin{array}{l}
p_{i,k,\mathsf{Hard}}-c_{i, k}-\lambda_{1}^{w_{i}} \mathcal{U}^{min}>\mathbbm{y}^{-}  \qquad\qquad\qquad\quad\quad~~\text{(33a)}\\
1-\frac{\Phi\left(\frac{p_{i, k, \mathsf{Hard}}-c_{i, k}-\lambda_{1}^{w_{i}} \mathcal{U}^{min}-\mu_{Y}}{\sigma_{Y}}\right)-\Phi\left(\frac{\mathbbm{y}^{-}-\mu_{Y}}{\sigma_{Y}}\right)}{\Phi\left(\frac{\mathbbm{y}^{+}-\mu_{Y}}{\sigma_{Y}}\right)-\Phi\left(\frac{\mathbbm{y}^{-}-\mu_{Y}}{\sigma_{Y}}\right)} \leq \lambda_{2}^{w_{i}}\text{(33b)}
\end{array}\right.\notag
\end{equation}
From (33a), we have (34),
\begin{align}
\label{eq34}
p_{i,k,\mathsf{Hard}}^*>\lambda_1^{w_i}\mathcal{U}^{min}+c_{i,k}+\xi_i r_i q_{i,k,\mathsf{Hard}} b_{w_i}^{-}\tag{34}
\end{align}
As for (33b), we have the following (35). Since error function erf(.) represents a non-elementary function which pose challenges to have a close form of $p_{i,k,\mathsf{Hard}}^*$. Thus, let $\widehat{p_{i,k,\mathsf{Hard}}^*}$ be the minimum tolerable value of payment $p_{i,k,\mathsf{Hard}}$ that meets inequation $\text{erf}\left(\frac{p_{i,k,\mathsf{Hard}}-c_{i,k}-\lambda_1^{w_i}\mathcal{U}^{min}-\mu_Y)}{\sqrt{2}\sigma_Y}\right)\geq 2\mathbbm{c}_i-1$ given by (35), for notational simplicity. Then, considering constraint C5 under $x_{i,k,\mathsf{Hard}}=1$, we have (36). In conclusion, for all $w_i\in \mathbb{W}$, $\bm{s_k}\in\mathbb{S}$, $p_{i,k,\mathsf{Hard}}^*$ can be defined by (37).

\vfill
\begin{strip}
\hrulefill
\begin{align}
&1-\frac{\Phi\left(\frac{p_{i,k,\mathsf{Hard}}-c_{i, k}-\lambda_{1}^{w_{i}}\mathcal{U}^{min }-\mu_{Y}}{\sigma_{Y}}\right)-\Phi\left(\frac{\mathbbm{y}^{-}-\mu_{Y}}{\sigma_{Y}}\right)}{\Phi\left(\frac{\mathbbm{y}^{+}-\mu_{Y}}{\sigma_{Y}}\right)-\Phi\left(\frac{\mathbbm{y}^{-}-\mu_{Y}}{\sigma_{Y}}\right)} \leq \lambda_{2}^{w_{i}} \xLongrightarrow {\mathbbm{c}_{i}\triangleq\left(1-\lambda_{2}^{w_{i}}\right)\left(\Phi\left(\frac{\mathbbm{y}^{+}-\mu_{Y}}{\sigma_{Y}}\right)-\Phi\left(\frac{\mathbbm{y}^{-}-\mu_{Y}}{\sigma_{Y}}\right)\right)+\Phi\left(\frac{\mathbbm{y}^{-}-\mu_{Y}}{\sigma_{Y}}\right)}\notag\\
&\Phi\left(\frac{p_{i,k,\mathsf{Hard}}-c_{i, k}-\lambda_{1}^{w_{i}} \mathcal{U}^{min}-\mu_{Y}}{\sigma_{Y}}\right) \geq \mathbbm{c}_{i} \Rightarrow \int_{-\infty}^{\frac{p_{i,k,\mathsf{Hard}}-c_{i, k}-\lambda_{1}^{w_{i}} \mathcal{U}^{min}-\mu_{Y}}{\sigma_{Y}}} \exp \left(-\frac{t^{2}}{2}\right) dt \geq \sqrt{2 \pi} \mathbbm{c}_{i} \notag \\
&{\Rightarrow} \sqrt{\frac{\pi}{2}} \operatorname{erf}\left(\frac{p_{i,k,\mathsf{Hard}}-c_{i, k}-\lambda_{1}^{w_{i}} \mathcal{U}^{min}-\mu_{Y}}{\sqrt{2} \sigma_{Y}}\right)+\sqrt{\frac{\pi}{2}} \geq \sqrt{2 \pi} \mathbbm{c}_{i} \Rightarrow \operatorname{erf}\left(\frac{p_{i,k,\mathsf{Hard}}-c_{i,k}-\lambda_{1}^{w_{i}} \mathcal{U}^{min}-\mu_{Y}}{\sqrt{2} \sigma_{Y}}\right) \geq 2 \mathbbm{c}_{i}-1\tag{35}\\

 \label{eq36}
&\overline{\mathcal{U}^{w_i}}\left(x_{i,k,\mathsf{Hard}}=1, q_{i,k,\mathsf{Hard}},p_{i,k,\mathsf{Hard}},\alpha_i,\beta_i\right)=a_i(p_{i,k,\mathsf{Hard}}-c_{i,k,})-\xi_ir_ia_iq_{i,k,\mathsf{Hard}}\overline{\beta_i}\notag\\
&{\Longrightarrow}~p_{i,k,\mathsf{Hard}}>\xi_ir_iq_{i,k,\mathsf{Hard}}\overline{\beta_i}+c_{i,k}\tag{36}\\

 \label{eq37}
&p_{i,k,\mathsf{Hard}}^*=\text{max}\left(\lambda_1^{w_i}\mathcal{U}^{min}+c_{i,k}+\xi_ir_iq_{i,k,\mathsf{Hard}}b_{w_i}^{-}; ~\widehat{p_{i,k,\mathsf{Hard}}^*};~ \xi_ir_iq_{i,k,\mathsf{Hard}}\overline{\beta_i}+c_{i,k}\right)+\Delta p\tag{37}
\end{align}
\end{strip}

\section{Analysis of Constraints (C1) and (C7)}
\noindent
Notably, it is difficult to obtain exact tractable forms for $\mathcal{R}_1^{s_k} \left(\mathbb{X},\mathbb{Q},\mathbb{A}\right)$ and $\mathcal{R}_2^{s_k}\left(\mathbb{X},\mathbb{P}^*,\mathbb{A}\right)$ since they contain mixed random variables that are independent with different distributions. Let $\mathbb{I}_{k,\mathsf{Hard}}$ and $\mathbb{I}_{k,\mathsf{Soft}}$ denote the set of workers which are hired to offer hard assurance, and soft assurance of sensing service for task $\bm{s_k}$, respectively, for notational simplicity ($\mathbb{I}_{k,\mathsf{Hard}}\cap\mathbb{I}_{k,\mathsf{Soft}} =\emptyset$). For example, obtaining the value range of $\sum_{w_i\in\mathbb{W}}\sum_{\ell\in\bm{L}}\alpha_ix_{i,k,\ell}q_{i,k,\ell}$ in $\mathcal{R}_1^{s_k}(\mathbb{X},\mathbb{Q},\mathbb{A})$ needs considering $2+C_{|\mathbb{I}_{k,\ell'}|}^1+C_{|\mathbb{I}_{k,\ell'}|}^2+\cdots+C_{|\mathbb{I}_{k,\ell'}|}^{|\mathbb{I}_{k,\ell'}|-1}$ possible situations (where $\ell'=$~“Soft”); while (C7) needs considering $2+C_{\sum_{\ell\in\bm{L}}|\mathbb{I}_{k,\ell}|}^1+C_{\sum_{\ell\in\bm{L}}|\mathbb{I}_{k,\ell}|}^2+\cdots+C_{\sum_{\ell\in\bm{L}}|\mathbb{I}_{k,\ell}|}^{\sum_{\ell\in\bm{L}}|\mathbb{I}_{k,\ell}|-1}$ possible situations, which further pose challenges on analyzing risk-related constraints.

We first rewrite constraint (C1) as inequality (38). Then, to obtain a tractable expression of the $\operatorname{Pr}\{\}$ in (38), we obtain (39) according to Markov inequality~\cite{20}, where (C1) is turned into (40) based on (38) and (39). 
Similarly, to obtain a tractable expression, $\mathcal{R}_2^{s_k}(\mathbb{X},\mathbb{P}^*,\mathbb{A})$ is rewritten as the following (41), which uses the result of Markov inequality. Constraints (C7) can thus be obtained as (42).
\begin{strip}
\hrulefill
\begin{align}
\label{eq38}
&\mathcal{R}_{1}^{s_{k}}\left(\mathbb{X}, \mathbb{Q}, \mathbb{A}\right)\le \lambda_3^{s_k}\Longrightarrow \operatorname{Pr}\left\{{\sum_{w_i\in\mathbb{W}}\sum\limits_{\ell\in\bm{L}}\alpha_ix_{i,k,\ell}q_{i,k,\ell}}> \lambda_1^{s_k}d_k^{Q}\right\}\ge 1-\lambda_3^{s_k}
\tag{38}
\end{align}

\begin{align}
\label{39}
&\operatorname{Pr}\left\{{\sum_{w_i\in\mathbb{W}}\sum\limits_{\ell\in\bm{L}}\alpha_ix_{i,k,\ell}q_{i,k,\ell}}> \lambda_1^{s_k}d_k^{Q}\right\}
\le\operatorname{Pr}\left\{{\sum_{w_i\in\mathbb{W}}\sum\limits_{\ell\in\bm{L}}\alpha_ix_{i,k,\ell}q_{i,k,\ell}}\ge \lambda_1^{s_k}d_k^{Q}\right\}\notag\\
&\le \frac{\text{E}\left[\sum\limits_{w_i\in\mathbb{W}}\sum\limits_{\ell\in\bm{L}}\alpha_ix_{i,k,\ell}q_{i,k,\ell}\right]}{\lambda_1^{s_k}d_k^{Q}}= \frac{\sum\limits_{w_i\in\mathbb{W}}\sum\limits_{\ell\in\bm{L}}a_ix_{i,k,\ell}\overline{q_{i,k,\ell}}}{\lambda_1^{s_k}d_k^{Q}}
\tag{39}
\end{align}

\begin{align}
\label{40}
&\text{(C1)}^\prime:~\frac{\sum\limits_{w_i\in\mathbb{W}}\sum\limits_{\ell\in\bm{L}}a_ix_{i,k,\ell}\overline{q_{i,k,\ell}}}{\lambda_1^{s_k}d_k^{Q}} \ge 1-\lambda_3^{s_k},\forall \bm{s_k}\in\mathbb{S}
\tag{40}
\end{align}

\begin{align}
\label{41}
&\mathcal{R}_2^{s_k}\left(\mathbb{X},\mathbb{P}^*,\mathbb{A}\right)=\operatorname{Pr}\left\{\sum\limits_{w_i\in\mathbb{W}}\sum\limits_{\ell\in\bm{L}}\alpha_ix_{i,k,\ell}p_{i,k,\ell}^*>\lambda_2^{s_k}d_k^{B}\right\}\le \operatorname{Pr}\left\{\sum_{w_i\in\mathbb{W}}\sum\limits_{\ell\in\bm{L}}\alpha_ix_{i,k,\ell}p_{i,k,\ell}^*\ge\lambda_2^{s_k}d_k^{B}\right\}\notag\\
&\le \frac{\text{E}\left[\sum\limits_{w_i\in\mathbb{W}}\sum\limits_{\ell\in\bm{L}}\alpha_ix_{i,k,\ell}p_{i,k,\ell}^*\right]}{\lambda_2^{s_k}d_k^{B}}=\frac{\sum\limits_{w_i\in\mathbb{W}}\sum\limits_{\ell\in\bm{L}}a_ix_{i,k,\ell}p_{i,k,\ell}^*}{\lambda_2^{s_k}d_k^{B}}\tag{41}
\end{align}

\begin{align}
\label{42}
&\text{(C7)}^\prime:~\frac{\sum\limits_{w_i\in\mathbb{W}}\sum\limits_{\ell\in\bm{L}}a_ix_{i,k,\ell}p_{i,k,\ell}^*}{\lambda_2^{s_k}d_k^{B}}\le\lambda_4^{s_k},\forall \bm{s_k}\in\mathbb{S}\tag{42}
\end{align}

\hrulefill
\end{strip}

\section{Analysis of (24)}

First, let variable $u_{i,k,\ell}\triangleq\ln {x_{i,k,\ell}}$, and $v\triangleq\ln y$; while $\mathbbm{u}={\{u_{i,k,\ell}\}}_{w_i\in \mathbb{W},\bm{s_k}\in \mathbb{S},\ell\in\bm{L}}$ for notational simplicity. 
By applying $z_{i,k,\ell}$ and $v$, $\bm{\mathcal{P}_3}$ is reformulated by a standard convex optimization problem, as the following (43). 
\begin{align}
&\min\limits_{\mathbb{X},y}y \Rightarrow \min\limits_{\mathbbm{u}, v} \text{e}^v \tag{43}
\end{align}
Constraints ${(\hat{\text{C}}\text{1})}^\prime, \text{(C5)},  {\text{(C7)}}^\prime, {(\hat{\text{C}}\text{8})}^\prime, (\hat{\text{C}}\text{9})$ can be transferred by the following (44)-(49), by applying the logarithmic change of variables. Specifically, similar to ${\bm{x}}^{[m]}$, let ${\mathbbm{u}}^{[m]}=\left\{u_{i,k,\ell}^{[m]}\right\}$, and $v^{[m]}$ denote the solution of $m$-th iteration. Based on the above discussions, problem $\bm{\mathcal{P}_3}$ is reformulated as $\bm{\mathcal{P}_5}$ given in (50), which is a typical convex optimization problem. 
\vspace{1cm}
\begin{align}
&{(\hat{\text{C}}\text{1})}^\prime: \forall \bm{s_k}\in\mathbb{S}\notag\\
&\frac{(1-\lambda_3^{s_k})\lambda_1^{s_k}d_k^{Q}}
{\prod\limits_{w_i\in\mathbb{W}}\prod\limits_{\ell\in\bm{L}}{\left(\frac{\text{e}^{u_{i,k,\ell}}f_k^{(C1)}{\left(\text{e}^{\mathbbm{u}^{[m-1]}}\right)}}{\text{e}^{u_{i,k,\ell}^{[m-1]}}}\right)}
^{\frac{a_i\overline{q_{i,k,\ell}}\text{e}^{u_{i,k,\ell}^{[m-1]}}}{f_k^{(C1)}\left(\text{e}^{\mathbbm{u}^{[m-1]}}\right)}}}\le 1,\tag{44}\\
&\text{(C4)}^\prime:u_{i,k,\ell}\le 0, \forall w_i\in\mathbb{W},\bm{s_k}\in\mathbb{S}, \ell\in\bm{L}\tag{45}\\
&\text{(C5)}:\sum_{\bm{s_k}\in\mathbb{S}}\sum_{\ell\in\bm{L}}\text{e}^{u_{i,k,\ell}}\le 1,\forall w_i\in \mathbb{W}\tag{46}\\
&{\text{(C7)}}^\prime:\frac{{\sum\limits_{w_i\in\mathbb{W}}\sum\limits_{\ell\in\bm{L}}a_ip_{i,k,\ell}^*\text{e}^{u_{i,k,\ell}}}}{\lambda_2^{s_k}\lambda_4^{s_k}d_k^{B}}\le 1,\forall \bm{s_k}\in\mathbb{S}\tag{47}
\end{align}
\clearpage
\newpage

\begin{strip}
\hrulefill
\begin{align}
&{(\hat{\text{C}}\text{8})}^\prime:\frac{\sum\limits_{w_i\in\mathbb{W}}\sum\limits_{\bm{s_k}\in\mathbb{S}}\sum\limits_{\ell\in\bm{L}}\text{e}^{u_{i,k,\ell}}}
{\prod\limits_{w_i\in\mathbb{W}}\prod\limits_{\bm{s_k}\in\mathbb{S}}\prod\limits_{\ell\in\bm{L}}{\left(\frac{\left(\mu^\prime+\text{e}^{2u_{i,k,\ell}}\right)f^{(C8)}\left(\text{e}^{\mathbbm{u}^{[m-1]}}\right)}{\mu^\prime+\text{e}^{2u_{i,k,\ell}^{[m-1]}}}\right)}
^{\frac{
{\mu^\prime+\text{e}^{2u_{i,k,\ell}^{[m-1]}}}}{f^{(C8)}\left(\text{e}^{\mathbbm{u}^{[m-1]}}\right)}}}\le 1\tag{48}\\
&(\hat{\text{C}}\text{9}): \frac{1}{\prod\limits_{w_i\in\mathbb{W}}\prod\limits_{\bm{s_k}\in\mathbb{S}}\prod\limits_{\ell\in\bm{L}}{\left(\frac{\text{e}^{(v+u_{i,k,\ell})}f^{(C9)}\left(\text{e}^{\mathbbm{u}^{[m-1]}},\text{e}^{v^{[m-1]}}\right)}{\text{e}^{\left(v^{[m-1]}+u_{i,k,\ell}^{[m-1]}\right)}}\right)}^{\frac{a_i\overline{q_{i,k,\ell}}\text{e}^{\left(v^{[m-1]}+u_{i,k,\ell}^{[m-1]}\right)}}{f^{(C9)}\left(\text{e}^{\mathbbm{u}^{[m-1]}},\text{e}^{v^{[m-1]}}\right)}}}\le 1
\tag{49}
\end{align}

\begin{align*}
&\qquad\qquad\qquad\qquad\qquad\qquad\qquad\qquad\qquad\qquad\qquad\bm{\mathcal{P}_5:}
\min\limits_{\mathbbm{u},v}\text{e}^v\tag{50}\\
&\textrm{s.t.} ~\text{(C4)}^\prime, \notag\\
&{(\hat{\text{C}}\text{1})}^\prime: \ln\left({(1-\lambda_3^{s_k})\lambda_1^{s_k}d_k^{Q}}\right)-
\sum\limits_{w_i\in\mathbb{W}}\sum\limits_{\ell\in\bm{L}}\ln\left({\left(\frac{f_k^{(C1)}\left(\text{e}^{\mathbbm{u}^{[m-1]}}\right)}{\text{e}^{u_{i,k,\ell}^{[m-1]}}}\right)}^{\frac{a_i\overline{q_{i,k,\ell}}\text{e}^{u_{i,k,\ell}^{[m-1]}}}{f_k^{(C1)}\left(\text{e}^{\mathbbm{u}^{[m-1]}}\right)}}
\right)\notag\\
&-\sum\limits_{w_i\in\mathbb{W}}\sum\limits_{\ell\in\bm{L}}{\frac{a_i\overline{q_{i,k,\ell}}\text{e}^{u_{i,k,\ell}^{[m-1]}}u_{i,k,\ell}}{f_k^{(C1)}\left(\text{e}^{\mathbbm{u}^{[m-1]}}\right)}}\le 0,  \forall \bm{s_k}\in\mathbb{S}\\
&\text{(C5)}: \ln\left(\sum\limits_{\bm{s_k}\in\mathbb{S}}\sum\limits_{\ell\in\bm{L}}\text{e}^{u_{i,k,\ell}}\right)\le 0, \forall w_i\in \mathbb{W}\\
&{\text{(C7)}}^\prime:\ln\left(\frac{{\sum\limits_{w_i\in\mathbb{W}}\sum\limits_{\ell\in\bm{L}}a_ip_{i,k,\ell}^*\text{e}^{u_{i,k,\ell}}}}{\lambda_2^{s_k}\lambda_4^{s_k}d_k^{B}}\right)\le 0 \Rightarrow
 \ln\left(\sum\limits_{w_i\in\mathbb{W}}\sum\limits_{\ell\in\bm{L}}a_ip_{i,k,\ell}^*\text{e}^{u_{i,k,\ell}}\right)-\ln\left(\lambda_2^{s_k}\lambda_4^{s_k}d_k^{B}\right)\le 0, \forall \bm{s_k}\in\mathbb{S}\\

&{(\hat{\text{C}}\text{8})}^\prime: \ln\left(\sum\limits_{w_i\in\mathbb{W}}\sum\limits_{\bm{s_k}\in\mathbb{S}}\sum\limits_{\ell\in\bm{L}}\text{e}^{u_{i,k,\ell}}\right)-\sum\limits_{w_i\in\mathbb{W}}\sum\limits_{\bm{s_k}\in\mathbb{S}}\sum\limits_{\ell\in\bm{L}}\ln\left({\left(\frac{f^{(C8)}\left(\text{e}^{\mathbbm{u}^{[m-1]}}\right)}{\mu^\prime+\text{e}^{2u_{i,k,\ell}^{[m-1]}}}\right)}
^{\frac{
{\mu^\prime+\text{e}^{2u_{i,k,\ell}^{[m-1]}}}}{f^{(C8)}\left(\text{e}^{\mathbbm{u}^{[m-1]}}\right)}}
\right)-\notag\\
&\sum\limits_{w_i\in\mathbb{W}}\sum\limits_{\bm{s_k}\in\mathbb{S}}\sum\limits_{\ell\in\bm{L}}\ln\left({\left(\mu^\prime+\text{e}^{2u_{i,k,\ell}}\right)}^{\frac{
{\mu^\prime+\text{e}^{2u_{i,k,\ell}^{[m-1]}}}}{f^{(C8)}\left(\text{e}^{\mathbbm{u}^{[m-1]}}\right)}}\right)\le 0 \notag\\
&(\hat{\text{C}}\text{9}): -\sum\limits_{w_i\in\mathbb{W}}\sum\limits_{\bm{s_k}\in\mathbb{S}}\sum\limits_{\ell \in\bm{L}}\ln\left({\left(\frac{f^{(C9)}\left(\text{e}^{\mathbbm{u}^{[m-1]}},\text{e}^{v^{[m-1]}}\right)}{\text{e}^{\left(v^{[m-1]}+u_{i,k,\ell}^{[m-1]}\right)}}\right)}^{{\frac{a_i\overline{q_{i,k,\ell}}\text{e}^{\left(v^{[m-1]}+u_{i,k,\ell}^{[m-1]}\right)}}{f^{(C9)}\left(\text{e}^{\mathbbm{u}^{[m-1]}},\text{e}^{v^{[m-1]}}\right)}}}\right) \notag\\
&-\sum\limits_{w_i\in\mathbb{W}}\sum\limits_{\bm{s_k}\in\mathbb{S}}\sum\limits_{\ell \in\bm{L}}\frac{\left(v+u_{i,k,\ell}\right)a_i\overline{q_{i,k,\ell}}\text{e}^{\left(v^{[m-1]}+u_{i,k,\ell}^{[m-1]}\right)}}{f^{(C9)}\left(\text{e}^{\mathbbm{u}^{[m-1]}},\text{e}^{v^{[m-1]}}\right)}\le 0
\end{align*}

\hrulefill
\end{strip}

\section{Analysis of (25)}
\noindent
The standard GP format of temporary worker recruitment $\bm{\mathcal{P}_4}$ is shown by the following $\bm{\mathcal{P}_6}$.
\begin{align}
\bm{\mathcal{P}_6:}
\min\limits_{\mathbb{X}^\prime,y^\prime}~y^\prime \tag{51}
 \end{align}
\begin{align}
&\textrm{s.t.}~\text{(C11)},\notag\\
&\text{(C12)}: \frac{\sum\limits_{w_{i^\prime}\in\mathbb{W}^\prime}\sum\limits_{\ell\in\bm{L}}x_{i^\prime, k^\prime, \ell}p_{i^\prime, k^\prime, \ell}}{\left(d^{B}_{k^\prime}\right)^\prime}\le 1, \forall \bm{s_{k^\prime}}\in\mathbb{S}^\prime\notag\\
&\text{(C13)}: 0\le x_{i^\prime, k^\prime, \ell}\le 1, \forall w_{i^\prime}\in\mathbb{W}^\prime, \bm{s_{k^\prime}}\in \mathbb{S}^\prime,\ell \in \bm{L}\notag\\
&\text{(C14)}: \frac{\sum\limits_{w_{i^\prime}\in\mathbb{W}^\prime}\sum\limits_{\bm{s_{k^\prime}}\in \mathbb{S}^\prime}\sum\limits_{\ell\in\bm{L}}x_{i^\prime, k^\prime, \ell}}{\mu+\sum\limits_{w_{i^\prime}\in\mathbb{W}^\prime}\sum\limits_{\bm{s_{k^\prime}}\in \mathbb{S}^\prime}\sum\limits_{\ell\in\bm{L}}\left(x_{i^\prime, k^\prime, \ell}\right)^2}\le 1\notag\\
&\text{(C15)}: \frac{1}{y^\prime\sum\limits_{w_{i^\prime}\in\mathbb{W}^\prime}\sum\limits_{\bm{s_{k^\prime}}\in \mathbb{S}^\prime}\sum\limits_{\ell\in\bm{L}}x_{i^\prime, k^\prime, \ell}q_{i^\prime, k^\prime, \ell}}\le 1\notag
\end{align}
Let $\bm{x}^\prime=\{x_{i^\prime, k^\prime, \ell}\}$, while $\bm{x}^{\prime[m]}$, $x_{i^\prime,k^\prime,\ell}^{[m]}$ and $y^{\prime{[m]}}$ denote the solution of the $m$-th iteration. Besides, let $f^{(C14)}$ and $f^{(C15)}$ be the denominator of the left side of inequality in constraints (C14) and (C15), where the approximations are shown by (52) and (53). Thus, constraints (C14) and (C15) can be rewritten by ${(\hat{\text{C}}\text{14})}$ and ${(\hat{\text{C}}\text{15})}$. 
By applying logarithmic change of variables, the standard formed GP is transformed into a convex optimization problem $\bm{\mathcal{P}_7}$ which can further be solved by optimization tools offered by Matlab. Specifically,  $x_{i^\prime, k^\prime, \ell}\triangleq \text{e}^{u_{i^\prime,k^\prime,\ell}}$, $y^\prime \triangleq \text{e}^{v^\prime}$, $\mathbbm{u}^\prime={\{u_{i^\prime,k^\prime,\ell}\}}_{w_{i^\prime}\in\mathbb{W}^\prime, \bm{s_{k^\prime}}\in\mathbb{S}^\prime,\ell\in\bm{L}}$. 


\begin{strip}
\hrulefill
\begin{align}
& f^{(C14)}\left(\bm{x}^\prime\right)=\mu+\sum\limits_{w_{i^\prime}\in\mathbb{W}^\prime}\sum\limits_{\bm{s_{k^\prime}}\in \mathbb{S}^\prime}\sum\limits_{\ell\in\bm{L}}\left(x_{i^\prime, k^\prime, \ell}\right)^2 
\Rightarrow f^{(C14)}(\bm{x}^\prime)\ge \hat{f}^{(C14)}(\bm{x}^\prime)\notag\\
&\triangleq \prod\limits_{w_{i^\prime}\in\mathbb{W}^\prime}\prod\limits_{\bm{s_{k^\prime}}\in\mathbb{S}^\prime}\prod\limits_{\ell\in\bm{L}}{\left(\frac{\left(\mu^\prime+{(x_{i^\prime,k^\prime,\ell})}^2\right)f^{(C14)}\left(\bm{x}^{\prime[m-1]}\right)}{\mu^\prime+\left({x_{i^\prime,k^\prime,\ell}^{[m-1]}}\right)^{2}}\right)}^{\frac{\mu^\prime+\left({x_{i^\prime,k^\prime,\ell}^{[m-1]}}\right)^{2}}{f^{(C14)}\left(\bm{x}^{\prime[m-1]}\right)}}
\tag{52}\\
& f^{(C15)}\left(\bm{x}^\prime, y^\prime\right)=\sum\limits_{w_{i^\prime}\in\mathbb{W}^\prime}\sum\limits_{\bm{s_{k^\prime}}\in\mathbb{S}^\prime}\sum\limits_{\ell\in\bm{L}} y^\prime q_{i^\prime,k^\prime,\ell}x_{i^\prime,k^\prime,\ell}\Rightarrow f^{(C15)}(\bm{x}^\prime,y^\prime) \ge \hat{f}^{(C15)}(\bm{x}^\prime,y^\prime)\notag\\
&\triangleq \prod\limits_{w_{i^\prime}\in\mathbb{W}^\prime}\prod\limits_{\bm{s_{k^\prime}}\in\mathbb{S}^\prime}\prod\limits_{\ell\in\bm{L}}{\left(\frac{x_{i^\prime,k^\prime,\ell}f^{(C15)}\left(\bm{x}^{\prime [m-1]},y^{\prime[m-1]}\right)}{y^{\prime[m-1]}x_{i^\prime,k^\prime,\ell}^{[m-1]}}\right)}^{\frac{q_{i^\prime,k^\prime,\ell}y^{\prime [m-1]}x_{i^\prime,k^\prime,\ell}^{[m-1]}}{f^{(C15)}\left(\bm{x}^{\prime [m-1]},y^{\prime[m-1]}\right)}}
\tag{53}\\
&{(\hat{\text{C}}\text{14})}: \frac{\sum\limits_{w_{i^\prime}\in\mathbb{W}^\prime}\sum\limits_{\bm{s_{k^\prime}}\in \mathbb{S}^\prime}\sum\limits_{\ell\in\bm{L}}x_{i^\prime, k^\prime, \ell}}{f^{(C14)}}\le 1 \tag{54}\\
&{(\hat{\text{C}}\text{15})}: \frac{1}{f^{(C15)}}\le 1 \tag{55}
\end{align}
\begin{align}
&\qquad\qquad\qquad\qquad\qquad\qquad\qquad\qquad\qquad\qquad\qquad\bm{\mathcal{P}_7:}
\min\limits_{\mathbbm{u}^\prime,v^\prime}\text{e}^{v^\prime}\tag{56}\\
&\textrm{s.t.} \notag\\
&\text{(C11)}^\prime: \ln\left(\sum\limits_{\bm{s_{k^\prime}}\in\mathbb{S}^\prime}\sum\limits_{\ell\in\bm{L}}\text{e}^{u_{i^\prime,k^\prime,\ell}}\right)\le 0, \forall w_{i^\prime}\in \mathbb{W}^\prime,~~\text{(C12)}^\prime: \frac{\sum\limits_{w_{i^\prime}\in\mathbb{W}^\prime}\sum\limits_{\ell\in\bm{L}}\text{e}^{u_{i^\prime, k^\prime, \ell}}p_{i^\prime, k^\prime, \ell}}{\left(d^{B}_{k^\prime}\right)^\prime}\le 1, \forall \bm{s_{k^\prime}}\in\mathbb{S}^\prime\notag\\
&\text{(C13)}^\prime: {u_{i^\prime, k^\prime, \ell}}\le 0, \forall w_{i^\prime}\in\mathbb{W}^\prime, \bm{s_{k^\prime}}\in \mathbb{S}^\prime,\ell \in \bm{L}\notag\\
&{\text{(}\hat{\text{C}}\text{14)}}^\prime: \ln\left(\sum\limits_{w_{i^\prime}\in\mathbb{W}^\prime}\sum\limits_{\bm{s_{k^\prime}}\in\mathbb{S}^\prime}\sum\limits_{\ell\in\bm{L}}\text{e}^{u_{i^\prime,k^\prime,\ell}}\right)-\sum\limits_{w_{i^\prime}\in\mathbb{W}^\prime}\sum\limits_{\bm{s_{k^\prime}}\in\mathbb{S}^\prime}\sum\limits_{\ell\in\bm{L}}\ln\left({\left(\frac{f^{(C14)}\left(\text{e}^{\mathbbm{u}^{\prime[m-1]}}\right)}{\mu^\prime+\text{e}^{2u_{i^\prime,k^\prime,\ell}^{[m-1]}}}\right)}
^{\frac{
{\mu^\prime+\text{e}^{2u_{i^\prime,k^\prime,\ell}^{[m-1]}}}}{f^{(C14)}\left(\text{e}^{\mathbbm{u}^{\prime[m-1]}}\right)}}
\right)-\notag\\
&\sum\limits_{w_{i^\prime}\in\mathbb{W}^\prime}\sum\limits_{\bm{s_{k^\prime}}\in\mathbb{S}^\prime}\sum\limits_{\ell\in\bm{L}}\ln\left({\left(\mu^\prime+\text{e}^{2u_{i^\prime,k^\prime,\ell}}\right)}^{\frac{
{\mu^\prime+\text{e}^{2u_{i^\prime,k^\prime,\ell}^{[m-1]}}}}{f^{(C14)}\left(\text{e}^{\mathbbm{u}^{\prime[m-1]}}\right)}}\right)\le 0 \notag\\
&{\text{(}\hat{\text{C}}\text{15)}}^\prime: -\sum\limits_{w_{i^\prime}\in\mathbb{W}^\prime}\sum\limits_{\bm{s_{k^\prime}}\in\mathbb{S}^\prime}\sum\limits_{\ell \in\bm{L}}\ln\left({\left(\frac{f^{(C15)}\left(\text{e}^{\mathbbm{u}^{\prime[m-1]}},\text{e}^{v^{\prime[m-1]}}\right)}{\text{e}^{\left(v^{\prime[m-1]}+u_{i^\prime,k^\prime,\ell}^{[m-1]}\right)}}\right)}^{{\frac{q_{i^\prime,k^\prime,\ell}\text{e}^{\left(v^{\prime[m-1]}+u_{i^\prime,k^\prime,\ell}^{[m-1]}\right)}}{f^{(C15)}\left(\text{e}^{\mathbbm{u}^{\prime[m-1]}},\text{e}^{v^{\prime[m-1]}}\right)}}}\right)\notag\\
& -\sum\limits_{w_{i^\prime}\in\mathbb{W}^\prime}\sum\limits_{\bm{s_{k^\prime}}\in\mathbb{S}^\prime}\sum\limits_{\ell \in\bm{L}}\frac{\left(v^\prime+u_{i^\prime,k^\prime,\ell}\right)q_{i^\prime,k^\prime,\ell}\text{e}^{\left(v^{\prime[m-1]}+u_{i^\prime,k^\prime,\ell}^{[m-1]}\right)}}{f^{(C15)}\left(\text{e}^{\mathbbm{u}^{\prime[m-1]}},\text{e}^{v^{\prime[m-1]}}\right)}\le 0\notag
 \end{align}
\end{strip}
\section{Key notations}
Table 1 summarizes key notations used in this paper.
\begin{table}[h!t]
{\footnotesize
\caption{Major Notations (we omit $j$ from Section 3)}
\vspace{-3mm}
\begin{tabular}{|ll|}
\hline\\[-2.9mm]\hline
\multicolumn{1}{|c|} {\textbf{Notation} } &{ \textbf{Description} }\\
\hline
\multicolumn{1}{|c|} {$\mathbb{S}_j, \mathbb{W}$} & {Set of tasks, set of workers}\\
\hline
\multicolumn{1}{|c|} {$\bm{s_{j,k}}, d^B_{j,k}, d^Q_{j,k}$} & {MCSC task, its tolerable budget, its desired service quality} \\
\hline
\multicolumn{1}{|c|} {$w_i, \alpha_i, \beta_i$} & {MCSC worker, participation of $w_i$, local workload of $w_i$}\\
\hline
\multicolumn{1}{|c|} {$a_i, \mu_{w_i}, {(\sigma_{w_i})}^2, b^-_{w_i}, b^+_{w_i}$} &{Distribution parameters of $\alpha_i$ and $\beta_i$}\\
\hline
\multicolumn{1}{|c|} {$c_{i,j,k}, q_{i,j,k,\mathsf{Hard}}, q_{i,j,k,\mathsf{Soft}}$} &{Fundamental cost of $w_i$ for serving $\bm{s_{j,k}}$, hard, and soft service quality assurance}\\
\hline
\multicolumn{1}{|c|} {$p_{i,j,k,\mathsf{Hard}}, p_{i,j,k,\mathsf{Soft}}$} & {Service prices for hard and soft quality assurance}\\
\hline
\multicolumn{1}{|c|} {$\bm{L}, \ell$} & {The set of labels for hard/soft service quality assurance, and the corresponding index}\\ 
\hline
\multicolumn{1}{|c|} {$r_i, r_i^\prime$} & {Cost factor and the marginal performance degradation rate of $w_i$}\\ 
\hline
\multicolumn{1}{|c|} {$x_{i,k,\ell}$} & {Indicator of task-worker-service quality mapping} \\
\hline
\multicolumn{1}{|c|} {$\mathbb{P}, \mathbb{Q}, \mathbb{X}, \mathbb{A}, \mathbb{B}$} & {Profile of $p_{i,k,\ell}, q_{i,k,\ell}, x_{i,k,\ell}, \alpha_{i}, \beta_i$}\\
\hline
\multicolumn{1}{|c|} {$\mathcal{U}^{\mathbb{S}}, \overline{\mathcal{U}^{\mathbb{S}}}$} & {The overall utility of tasks, and its expectation}\\
\hline
\multicolumn{1}{|c|} {$\mathcal{U}^{w_i}, \overline{\mathcal{U}^{w_i}}$} & {The overall utility of worker $w_i$, and its expectation}\\
\hline
\multicolumn{1}{|c|} {$\mathcal{R}_1^{s_k},\mathcal{R}_2^{s_k},\mathcal{R}^{w_i}$} & {Risk of tasks and workers}\\
\hline
\multicolumn{1}{|c|} {$\lambda_1^{s_k},\lambda_2^{s_k}, \lambda_3^{s_k}, \lambda_4^{s_k}$} & {Threshold coefficient regarding risk analysis of tasks}\\
\hline
\multicolumn{1}{|c|} {$\lambda_1^{w_i},\lambda_2^{w_i}$} & {Threshold coefficient regarding risk analysis of workers}\\
\hline
\end{tabular}
\label{tab1}
}
\end{table}
\end{document}